\documentclass[nofootinbib,floatfix,superscriptaddress]{revtex4}        
\usepackage{graphicx}
\usepackage{epsfig}
\usepackage{bm}

\usepackage[T1]{fontenc}
\usepackage[latin9]{inputenc}
\usepackage{amssymb}
\usepackage{float}
\usepackage{amsmath}
\usepackage{dcolumn}
\usepackage{cancel}
\usepackage[colorlinks]{hyperref}
\usepackage[usenames,dvipsnames]{color}
\hypersetup{
     breaklinks=true,
    pdfstartview={FitH},    
    colorlinks=true,       
    linkcolor=blue,          
    citecolor=red,        
    filecolor=magenta,      
    urlcolor=blue,           
    anchorcolor=green,      
    linktocpage=true
}

\newcommand{\be}{\begin{equation}}
\newcommand{\ee}{\end{equation}}

\begin{document}

\title{Quantum Entropy-Driven Modifications to Holographic Dark Energy in $f(G,T)$ Gravity}

\author{Giuseppe Gaetano Luciano}
\email{giuseppegaetano.luciano@udl.cat}
\affiliation{Department of Chemistry, Physics and Environmental and Soil Sciences, Escola Politecninca Superior, Universidad de Lleida, Av. Jaume
II, 69, 25001 Lleida, Spain}

\date{\today}

\begin{abstract}
In this work, we present a \( f(G,T) \) gravity-based reconstruction of Barrow Holographic Dark Energy (BHDE). This approach extends the conventional HDE model by replacing the standard Bekenstein-Hawking entropy with Barrow entropy, which encapsulates quantum gravitational corrections to the geometry of black hole horizons. We explore the cosmological dynamics of a spatially flat Friedmann-Robertson-Walker background filled with a pressureless dust fluid, considering both conserved and non-conserved energy-momentum tensor models.
To this end, we employ the Hubble horizon as the infrared cutoff and adopt a power-law ansatz for the scale factor. We then investigate the evolution of key cosmological parameters, including the equation-of-state parameter \( \omega_{GT} \), the deceleration parameter \( q \), and the squared sound speed \( v_s^2 \).
Furthermore, we explore the dynamical behavior in the \( \omega_{GT} \)-\( \omega'_{GT} \) phase space. In the case of conserved energy-momentum tensor, our findings indicate that the BHDE model evolves from a quintessence-like regime into the phantom domain. This transition supports the current accelerated expansion of the Universe and offers an improvement over the original HDE model, which does not adequately account for the observed phenomenology. The corresponding \( \omega_{GT} \)-\( \omega'_{GT} \) trajectory lies within the freezing region of the phase space.
On the other hand, within the non-conserved framework, the BHDE model exhibits phantom-like behavior in the early Universe, subsequently evolving toward either a cosmological constant-like state or a quintessence-like regime. Notably, unlike the conserved scenario, the squared sound speed \( v_s^2 \) asymptotically attains positive values in the far future, signifying a stable configuration. Moreover, the trajectory in the \( \omega_{GT} \)-\( \omega'_{GT} \) phase space displays a thawing behavior. Finally, we evaluate the observational viability of our results and compare them with predictions from alternative reconstructed dark energy models.

\end{abstract}

 \maketitle

\section{Introduction}
\label{Intro}
The observed accelerated expansion of the present Universe~\cite{Supern,Supernbis,Supernter,Supernquar,Supernquin,ER} remains one of the most profound challenges in modern cosmology. Despite significant efforts, a definitive explanation for the origin of this phenomenon has yet to be established. Broadly, two main approaches have been proposed to address this issue. The first involves modifications to the geometric sector of the Einstein-Hilbert action, leading to a variety of frameworks collectively known as Extended Theories of Gravity~\cite{CapozDela}. The second approach introduces additional degrees of freedom in the matter sector, giving rise to dynamical dark energy (DE) models.
Within this latter category, a particularly promising candidate is the so-called Holographic Dark Energy (HDE) scenario~\cite{Cohen:1998zx,Horava:2000tb,Thomas:2002pq,Li:2004rb,Hsu:2004ri,Huang:2004ai,Wang:2005ph,Setare:2006sv,Guberina:2006qh,Granda,Sheykhi:2011cn,Bamba:2012cp,Ghaffari:2014pxa,Wang:2016och,Moradpour:2020dfm,Zhang,Li,Zhang2,Lu,Cap1}, which is grounded in the application of the holographic principle at cosmological scales.
More recently, hybrid frameworks have been developed by embedding HDE into various modified gravity theories, including \( f(R) \)~\cite{KK}, \( f(T) \)~\cite{teleparallel}, \( f(R,T) \)~\cite{Piattella,frt} and \( f(G) \)~\cite{fg} gravity, among others (here, \( R \), $G$ and $T$ denote the Ricci scalar, the Gauss-Bonnet invariant and the trace of the energy-momentum tensor, respectively). 

In its original formulation, HDE relies on the use of Bekenstein-Hawking entropy for the horizon degrees of freedom of the Universe, and adopts the Hubble horizon as the infrared (IR) cutoff. However, as shown in~\cite{Horava:2000tb,Thomas:2002pq,Li:2004rb}, this framework fails to reproduce the observed phenomenology of the Universe, thereby necessitating appropriate modifications.
Initial efforts to address this issue have explored the adoption of alternative IR cutoffs or the introduction of interactions between the dark sectors of the Universe (see~\cite{Wang:2016och,Wang:2016lxa} for a review). Recently, motivated by statistical considerations, various alternative models have been proposed based on generalized entropy formalisms, including Tsallis~\cite{Tsallis1,Tsallis1.5,Tsallis2,Tsallis3,Tsallis4,Tsallis5}, Kaniadakis~\cite{Kana1,Kana2,Kana3,Kana4} and Barrow~\cite{Bar1,Bar2,Bar3,Bar4,Bar5,Bar6,Bar7,Bar7.5,Bar8,Bar9} entropies.
It is important to note that, although these models may exhibit certain mathematical similarities, their underlying physical principles differ significantly. Specifically, they are inspired by nonextensive statistical mechanics (Tsallis), relativistic statistical frameworks (Kaniadakis) and quantum gravitational considerations (Barrow), respectively. In particular, Barrow entropy arises from the idea that quantum gravitational effects may deform the geometry of the horizon at small scales, leading to a fractal-like structure in spacetime~\cite{B2020}. This deviation from standard Bekenstein-Hawking entropy offers a phenomenological window into possible quantum corrections to gravitational dynamics.

While many extended HDE models based on entropy deformations can be tuned to achieve empirical consistency 
through specific choices of deformation parameters, 
their lack of an underlying Lagrangian formulation raises questions about why they should be regarded 
as fundamentally well-founded theories.
A promising strategy to overcome this limitation consists in employing \emph{reconstruction} methods~\cite{Rec1,Zhang2,Rec2}. This approach involves comparing the relative energy densities of extended HDE models and modified gravity theories to obtain an effective Lagrangian capable of reproducing the entire cosmic evolution. For instance, Tsallis HDE has been extensively studied within the frameworks of 
\( f(R) \)~\cite{EPL}, \( f(R,T) \)~\cite{Chinese}, \( f(G,T) \)~\cite{fgtgrav}, teleparallel~\cite{Wahe}, Brans--Dicke~\cite{Ghaffari:2018wks} and tachyon field~\cite{Liu:2021heo} gravity, 
yielding a rich and diverse cosmological phenomenology. More recently, analogous methodologies have been 
applied to the study of black hole physics~\cite{DAg}. By contrast, comparatively less attention has been 
devoted to the Barrow HDE (BHDE) framework, with only a limited number of theoretical contexts explored thus far~\cite{Sarkar,TeleBar,LucianoLiu,CosmLuc,BaRec1,BaRec2,BaRec3}. Nonetheless, this approach holds potential for offering valuable insights into the formulation of a consistent Lagrangian description of quantum gravity effects, owing to the fundamental motivations behind Barrow's conjecture~\cite{B2020}.

Motivated by these premises, in this paper we propose a reconstruction of BHDE within the framework of \( f(G,T) \) gravity~\cite{SHIK}, which has demonstrated strong potential in addressing the challenges posed by General Relativity in the context of quantum gravity~\cite{fgt1,fgt2,fgt3,fgt4}. This makes it a particularly suitable setting in which to investigate the effects induced by Barrow entropy.
As a specific background, we consider a spatially flat Friedmann-Robertson-Walker (FRW) Universe filled with a pressureless dust fluid. The cosmic evolution of this system is examined for both conserved and non-conserved energy-momentum tensor models, with particular emphasis on the observational viability of the resulting predictions. Furthermore, we compare our findings with those of other reconstructed dark energy models, emphasizing the distinctive advantages of our framework in reproducing key observational features of cosmic evolution.

The remainder of this work is organized as follows. 
To establish the reconstruction paradigm connecting \( f(G,T) \) gravity and BHDE, 
Sec.~\ref{ba} presents the essential features of both frameworks. 
The effective reconstruction procedure is then carried out in Sec.~\ref{reconef}. 
In Sec.~\ref{Devot}, we analyze the cosmological dynamics of the extended HDE model by evaluating 
the equation-of-state (EoS) parameter \( \omega_{GT} \), the deceleration parameter \( q \) 
and the squared sound speed \( v_s^2 \), as well as by examining the evolutionary trajectories 
in the \( \omega_{GT} \)-\( \omega'_{GT} \) phase space. In Sec.~\ref{obc}, we derive observational constraints on the model parameters using a dataset of 57 Hubble parameter measurements. Final remarks and future perspectives are presented in Sec.~\ref{Conc}. 
Additional mathematical details are provided in the Appendix \ref{AppMath}.
Throughout this work, we adopt natural units.

\section{Basics of $f(G,T)$ gravity and Barrow Holographic Dark Energy}
\label{ba}
Let us begin by reviewing \( f(G,T) \) gravity. To this end, we shall refer to the analysis presented in \cite{fgtgrav} for notation. 
The fundamental feature of \( f(G,T) \) gravity lies in the incorporation of curvature-matter coupling within the framework of modified Gauss-Bonnet theory, consistent with insights derived from string theory \cite{SHIK}. In this context, the gravitational Lagrangian is formulated by introducing an appropriate function \( f(G,T) \) into the Einstein-Hilbert action, thereby yielding
\be
\label{Action}
S=\int d^4x \left(-g\right)^{\frac{1}{2}} \left(\frac{R+f(G,T)}{2\kappa^2}+\mathcal{L}_m\right),
\ee
where $g$ and $\kappa$ are the determinant
of the metric tensor and the coupling
constant, respectively, while $\mathcal{L}_m$ is the
matter Lagrangian. To streamline the notation, we henceforth set \( \kappa = 1 \), and denote \( f(G,T) \) simply by \( f \), whenever no ambiguity arises.

The variation of the action~\eqref{Action} with respect to the metric tensor yields the corresponding field equations
\begin{eqnarray}
&&\hspace{0mm}R_{\alpha\beta}-\frac{R}{2}g_{\alpha\beta}\,=\, T_{\alpha\beta}+\frac{g_{\alpha\beta}}{2}f-\left(T_{\alpha\beta}+\Theta_{\alpha\beta}\right)f_T+\left(4R_{\chi\beta}R^\chi_\alpha-2RR_{\alpha\beta}+4R_{\alpha\chi\beta\eta}R^{\chi\eta}-2R_{\alpha}^{\chi\eta\gamma}R_{\beta\chi\eta\gamma}
\right)f_G\\[2mm]
\nonumber
&&\hspace{0mm}+\left(4R_{\alpha\beta}-2Rg_{\alpha\beta}\right)\Box f_G+2R\nabla_\alpha\nabla_\beta f_G-4R_\beta^\chi\nabla_\alpha\nabla_\chi f_G - 4 R_\alpha^\chi\nabla_\beta\nabla_\chi f_G+4g_{\alpha\beta} R^{\chi\eta}\nabla_\chi\nabla_\eta f_G-4R_{\alpha\chi\beta\eta}\nabla^\chi\nabla^\eta f_G\,,
\label{fe}
\end{eqnarray}
where $T_{\alpha\beta}$ is the energy-momentum tensor, 
$\Theta_{\alpha\beta}=g^{\sigma\xi}\left(\dfrac{\delta T_{\sigma\xi}}{\delta g^{\alpha\beta}}\right)$, $\nabla_\eta$ represents the covariant derivative
and $\Box=\nabla_\eta\nabla^\eta$ is the D'Alembert operator.
Furthermore, we have denoted by $f_G$ and $f_T$ the partial derivatives of $f$
with respect to $G$ and $T$, respectively.  

By taking the covariant divergence of the field equations, we derive the associated conservation equation
\be
\nabla^\alpha T_{\alpha\beta}\,=\,\frac{f_T}{1-f_T}\left[\left(\Theta_{\alpha\beta}+T_{\alpha\beta}\right)\nabla^\alpha\log f_T
-\frac{g_{\alpha\beta}}{2}\nabla^\alpha T+\nabla^\alpha\Theta_{\alpha\beta}\right].
\label{cons}
\ee
For a spatially flat Friedmann-Robertson-Walker (FRW) Universe with a perfect fluid matter configuration, the field equations take the form
\begin{subequations}
\label{Heq}
\begin{equation}
3H^2\,=\,\rho+\rho_{GT}\,\equiv\,\rho_{eff}\,,
  \end{equation}    
  \vspace{-6mm}
  \begin{equation}
-\left(2\dot H+3H^2\right)\,=\,p+p_{GT}\,\equiv\,p_{eff}\,,
 \end{equation}
\end{subequations}
where $H=\dot a/a$ is the Hubble parameter, $a$ is the time-dependent scale factor and we have defined the $f(G,T)$ gravity-induced corrections as
\begin{eqnarray}
\label{rho}
&&\hspace{-9mm}\rho_{GT}\,=\,\frac{f}{2}+\left(\rho+p\right)f_T-\frac{G}{2}f_G+12H^3\dot G f_{GG}+\,12H^3\dot T f_{GT}\,,\\[2mm]
&&\hspace{-9mm}p_{GT}\,=\,-\frac{f}{2}+\frac{G}{2}f_G-8H\left(\dot H+H^2\right)\left(\dot Gf_{GG}+\dot T f_{GT}\right)-4H^2\left(\dot G^2f_{GGG}+2\dot G\dot T f_{GGT}+\ddot T^2f_{GTT}+\ddot G f_{GG}+\ddot T f_{GT}
\right),
\end{eqnarray} 
with
\begin{eqnarray}
G&=&24H^2\left(H^2+\dot H\right),\\[2mm]
T&=&\rho-3p\,.
\label{T}
\end{eqnarray}
As per the usual convention, an overdot represents differentiation with respect to the cosmic time \( t \). It is straightforward to verify that Eqs.~\eqref{Heq} are consistent with the standard cosmological dynamics in the limiting case where our extended theory of gravity reduces to general relativity.

In the above setting, the conservation equation~\eqref{cons} for perfect fluid configuration can be expressed as 
\be
\label{cruc}
\dot\rho+3H(\rho+p)\,=\,-\frac{1}{1-f_T}\left[\left(\dot p+\frac{\dot T}{2}\right)f_T+\left(\rho+p\right)\dot f_T
\right].
\ee
Once the functional form of \( f(G,T) \) is specified, this equation serves as a key element in the subsequent reconstruction framework. Following~\cite{fgtgrav}, we consider the two specific forms given below:
\be
\label{f1}
f(G,T)\,=\,f_1(G)+f_2(T)\,\,\,\,\,\, (\mathrm{Model\,\,I})\,,\\[2mm]
\ee
which potentially involves minimal coupling between
curvature and matter contents of the Universe, 
ensuring that the interaction remains purely gravitational in nature. The chosen form of the generic function is not only manageable but also offers a clearer explanation of the present cosmic acceleration~\cite{Viable1,Viable2}. 

On the other hand, we shall consider
\be
f(G,T)\,=\,F(G) + \eta T\,,\,\,\,\,\,\,\, (\mathrm{Model\,\,II})\,,
\label{f2}
\ee
which simplifies to standard  $f(G)$ gravity in the limit as the arbitrary constant  $\eta \rightarrow 0$. Note that
the Model I typically corresponds to a configuration in which the energy-momentum tensor is conserved, provided a suitable choice of the function \( f_2(T) \) that ensures the vanishing of the covariant divergence, \( \nabla^\mu T_{\mu\nu} = 0 \). This constraint is nontrivial and imposes specific functional dependencies between the gravitational Lagrangian and the trace \( T \) of the energy-momentum tensor. By contrast, Model II encompasses scenarios where the non-minimal matter-geometry coupling gives rise to a non-conserved energy-momentum tensor. In such cases, the extra force arising from the non-conservation term may lead to significant phenomenological consequences~\cite{SHIK}.

\subsection{Barrow Holographic Dark Energy}
\label{HBa}
The Holographic Dark Energy (HDE) paradigm is theoretically grounded in the holographic principle~\cite{thooft,Thorn,Sussk}, which emerges from considerations in quantum gravity. This principle asserts that the physical description of a spatial region - along with all physical quantities defined within it - is not determined by its volume, but rather by the degrees of freedom encoded on its lower-dimensional boundary.
When applied to cosmology, this principle implies that the number of degrees of freedom - and consequently the entropy - of the Universe scales with the area of its horizon rather than its volume. That is, \( S \sim A = 4\pi L^2 \), where \( L \) denotes a characteristic cosmological length scale.

In accordance with this prescription, the energy content of the Universe, including that of dark energy (DE), should be described using quantities defined solely on its boundary. Consequently, the construction of the DE density \( \rho_{DE} \) relies on two fundamental quantities: the reduced Planck mass \( M_p \) (assumed to be unity under our units convention), which naturally emerges as the fundamental mass scale in quantum gravity, and the cosmological length scale \( L \), which serves as the infrared (IR) cutoff and encodes the size of the observable Universe~\cite{Wang:2016och}. 
Based on these considerations, and invoking the holographic bound, the total energy within a region of size \( L \) must not exceed the limit prescribed by the Bekenstein-Hawking relation, implying  the inequality \( \rho_{DE} \, L^4 \leq S \). Saturating this bound yields the standard expression for holographic dark energy
\begin{equation}
\label{HB}
\rho_{DE} = C L^{-2}\,,
\end{equation}
where the model-dependent parameter \( C \) has dimensions \( [L]^{-2} \) and must be constrained observationally~\cite{Li:2004rb}.

Recently, significant attention has been devoted to generalizations of the  HDE framework arising from modifications to the underlying entropy-area relation. In particular, Barrow~\cite{B2020} proposed that quantum fluctuations may deform the black hole geometry in such a way that the entropy associated with the horizon receives a power-law correction, which encodes information about the intricate, potentially fractal nature of spacetime at the quantum level. This leads to a generalized entropy formula of the form
\begin{equation}
    S_{\Delta}\sim A^{1+\frac{\Delta}{2}}\, ,
\end{equation}
where \( \Delta \in [0,1] \) is the so-called Barrow exponent, which quantifies the degree of quantum-gravitational deformation. The case \( \Delta = 1 \) represents the maximal deviation from the standard entropy-area relation, which is recovered in the limit \( \Delta = 0 \). Moreover, in the regime of small \( \Delta \), the Barrow entropy can be expanded as $S_\Delta \propto A\left(1 + \frac{\Delta}{2} \log A \right) + \mathcal{O}(\Delta^2)$.
Interestingly, entropy expressions involving logarithmic corrections are a common outcome across a wide range of quantum gravity theories, including string theory~\cite{Log1}, loop quantum gravity~\cite{Log2}, the AdS/CFT correspondence~\cite{Carlip} and frameworks based on generalized uncertainty principles~\cite{Log3}.  Additionally, the logarithmic structure emerges naturally from computations of entanglement entropy in four-dimensional spacetime within the ultraviolet (UV) limit~\cite{UV1}. This suggests that such correction terms may represent a universal signature of quantum gravitational effects.
Therefore, although in the subsequent analysis the entropy deformation is interpreted within the framework of Barrow's formulation as originating from quantum fluctuations on the horizon surface, the underlying mechanisms and resulting conclusions are expected to exhibit broad applicability across a wide range of quantum gravity theories.

Although the Barrow model was initially proposed in the context of black hole theory, it has been widely applied to the geometry of the Universe's horizon, motivated by the deep connection between gravity and thermodynamics~\cite{Bar1,Bar2,Bar3,Bar4,Bar5,Bar6,Bar7,Bar7.5,Bar8,Bar9}.
When the Barrow entropy is used in place of the usual area law, the bound~\eqref{HB} becomes~\cite{Bar1}
\begin{equation}
\label{Barbo}
    \rho_\Delta L^4 \le S_{\Delta}\,\, \Longrightarrow\,\, \rho_\Delta = C L^{\Delta-2}\,,
\end{equation}
where \( C \) must now carry dimensions of \( [L]^{-2-\Delta} \) to ensure dimensional consistency. In what follows, we shall refer to the HDE model constructed using Barrow entropy as Barrow Holographic Dark Energy (BHDE).

A central feature of the HDE framework is the choice of the IR cutoff scale \( L \), which plays a crucial role in shaping the resulting cosmological dynamics. In its original formulation~\cite{Li:2004rb}, the Hubble horizon \( L = H^{-1} \) was adopted as the IR cutoff. Although this choice is conceptually simple and geometrically well-motivated, it has been shown to be phenomenologically inconsistent with current observations, as it fails to yield a sufficiently negative equation of state to account for the Universe's accelerated expansion~\cite{Hsu:2004ri}.
To address this shortcoming, various alternative definitions of \( L \) have been proposed. A prominent example is the future event horizon~\cite{Li:2004rb}, which successfully produces acceleration but introduces non-locality and a degree of circular reasoning in its formulation. Another widely discussed approach is based on the Ricci scalar curvature, leading to Ricci HDE models~\cite{Gao:2007ep}, which are more local in nature and can achieve observational consistency under suitable conditions.

Despite the known shortcomings of adopting \( L = H^{-1} \) in the standard HDE scenario, we retain this choice in the present analysis. We demonstrate that, within the BHDE framework, it becomes viable for suitable values of the model parameters. Indeed, the deformation introduced by Barrow entropy enriches the underlying phenomenology, allowing our model to successfully account for late-time cosmic acceleration. This constitutes a significant improvement over the conventional HDE formulation constrained by the standard area-law entropy.

Hence, by setting $L$ equal to the Hubble radius in Eq.~\eqref{Barbo}, we find 
\be
\label{rhode}
\rho_\Delta\,=\,CH^{2-\Delta}\,,
\ee 
which provides the key input for the reconstruction procedure carried out below.

\section{Reconstruction Models of BHDE in $f(G, T)$ gravity}
\label{reconef}

In what follows, we reconstruct the Barrow Holographic Dark Energy (BHDE) model within the framework of $f(G,T)$ gravity, employing the energy density given in Eq.~\eqref{rhode} and assuming a perfect fluid configuration with dust matter ($p=0$). The analysis is carried out separately for the two models introduced in Eqs.~\eqref{f1} and~\eqref{f2}, respectively.

\subsection{Model I}
\label{ModI}
For the function \( f(G, T) \) in Eq.~\eqref{f1}, the corresponding field equations are provided in Eq.~\eqref{Heq} with $p=0$ and~\cite{fgtgrav}
\begin{eqnarray}
\label{rhogt}
\rho_{GT}&=&\frac{f_1(G)+f_2(T)}{2}+\rho f_{2T}(T)-\frac{G}{2}f_{1G}(G)+12H^3\dot G f_{1GG}(G)\,,\\[3mm]
p_{GT}&=&-\frac{f_1(G)+f_2(T)}{2}+\frac{G}{2}f_{1G}(G)-4H\left[2\left(\dot H+H^2\right)\dot G+H\ddot G\right]f_{1GG}(G)-4H^2\dot G^2f_{1GGG}(G)\,,
\label{pGT}
\end{eqnarray}
In turn, Eq.~\eqref{cruc} turns out to be
\be
\dot\rho+3H\rho\,=\,-\frac{1}{1-f_{2T}(T)}\left[\frac{\dot T}{2}f_{2T}(T)+T\dot f_{2T}(T)\right].
\ee

In order for the Lagrangian~\eqref{f1} to be consistent with the conservation of the energy-momentum tensor,  the right-hand side of the above expression must be taken equal to zero, i.e.~\cite{fgtgrav}
\be
\label{cont1}
\dot\rho+3H\rho\,=\,0\,, 
\ee
along with the further constraint
\be
\label{constra}
f_{2T}(T)+2Tf_{2TT}(T)\,=\,0\,\,\Longrightarrow\,\, f_2(T)\,=\,\eta_1 T^{1/2}+\eta_2\,,
\ee
where $\eta_i$ ($i = 1,2$) are integration constants. 

At this stage, in order to proceed with the analytical evaluation, we follow~\cite{PL0,PL1,PL2,PL3,PL4} and adopt a power-law ansatz for the scale factor, expressed in the form
\be
\label{at}
a(t)\,=\,a_0\left(\tau-t\right)^m\,, \qquad\, m>0\,,
\ee
where $\tau > t$ denotes a finite future time (see below more details) and we set $a_0=1$.\footnote{The subscript $0$ indicates the present value of the quantities to which it refers.}
Some comments are in order here: first, 
although alternative forms of the scale factor have been considered in the literature (e.g.,~\cite{Pasqua1,Pasqua2}), many of these ultimately exhibit behavior that can be characterized, at least asymptotically or effectively, by power-law dynamics. 
Notably, the form~\eqref{at} encompasses a broad range of cosmological phenomena, enabling the description of various evolutionary phases of the Universe depending on the value of the parameter \( m \). Specifically, it corresponds to an accelerated expansion for \( m > 1 \), while indicating a decelerated era for \( 0 < m < 1 \). In particular, the case \( m = \frac{2}{3} \) represents a dust-dominated Universe, and \( m = \frac{1}{2} \) corresponds to a radiation-dominated phase.


Using Eq.~\eqref{at}, we derive explicit expressions for the Hubble parameter, the total energy density and the Gauss--Bonnet invariant as~\cite{fgtgrav}
\begin{equation}
\label{Hdi}
H=-\frac{m}{\tau-t}\,,\qquad 
\rho=\rho_0\left(\tau-t\right)^{-3m}\,, \qquad 
G=\frac{24m^3\left(m-1\right)}{\left(\tau-t\right)^4}\,.
\end{equation}
These relations clearly illustrate the central physical role played by the parameter $\tau$: it represents a finite-time future singularity. Specifically, while the scale factor \( a(t) \) itself vanishes as \( t \to \tau^- \), the Hubble parameter, its time derivative, the total energy density and higher-order curvature invariants (such as the Gauss-Bonnet scalar) all diverge. This behavior is characteristic of a \emph{Type III finite-time singularity}~\cite{Div}.
The inclusion of this singularity time is not arbitrary; rather, it reflects a well-motivated physical scenario within non-standard cosmologies, particularly in theories with phantom-like behavior or exotic matter-geometry couplings. It sets a natural time scale governing the evolution and signals the breakdown of the classical description near \( t \to \tau \), where quantum gravity effects or ultraviolet (UV) completions may become relevant.

Now, the reconstruction paradigm is based on the identification of the BHDE and \( f(G,T) \) energy densities, given by Eqs.~\eqref{rhode} and~\eqref{rhogt}, respectively. From the former, and by making use of Eq.~\eqref{Hdi}, we are led to the following expressions~\cite{fgtgrav} 
\begin{eqnarray}
\rho_{GT}=C\left(-\frac{m}{\tau-t}\right)^{2-\Delta}\,, \qquad p_{GT}=\frac{m\left(2-3m\right)}{\left(\tau-t\right)^2}\,.
\end{eqnarray}

By further implementing the constraint~\eqref{constra} on $f_2(T)$ in Eq.~\eqref{rhogt}, the differential equation in $f_1(G)$ resulting from the reconstruction scheme is given by
\be
12H^3\dot Gf_{1GG}(G)
-\frac{G}{2}f_{1G}(G)+\frac{f_1(G)}{2}+
\eta_1 T^{1/2}+\frac{\eta_2}{2}\,=\,CH^{2-\Delta}=C\left(-\frac{m}{\tau-t}\right)^{2-\Delta}\,,
\label{CH}
\ee
where we have used Eq.~\eqref{T}.

Using the ansatz~\eqref{at}, upon performing algebraic manipulations, we obtain the following expression for \( f_1(G) \):
\begin{eqnarray}
\label{secf}
&&\hspace{-2mm}f_1(G)\,=\,c_1G+c_2G^{\frac{1-m}{4}}+\frac{1}{12\left(m+3\right)}
\left\{54^{-\frac{(2+\Delta)}{4}}C\hspace{0.2mm}m\hspace{0.4mm}\left[-\frac{m}{6\left[m^3\left(m-1\right)\right]^{\frac{1}{4}}}\right]^{-(2+\Delta)}
\left[\frac{4\left(3+m\right)}{\left(1+m-\Delta\right)\left(2+\Delta\right)}
\right]G^{\frac{2-\Delta}{4}}\right.\\[2mm]
\nonumber
&&\hspace{-2mm}+\,24\left(\frac{2}{3}\right)^{\frac{m}{8}}\hspace{-1mm}\eta_1{\rho_0^{\frac{1}{2}}}\left(m-1\right)\left[m^3\left(m-1\right)\right]^{-\frac{3m}{8}}\left[2^{-m}\left(\frac{8}{3m-8}\right)-76^{-\frac{m}{4}}\left(\frac{8}{5m-2}\right)
\right]G^{\frac{3m}{8}}-12\eta_2\left(m-1\right)\left(\frac{m-5}{m-1}\right)
\Bigg\},
\label{f1Gne}
\end{eqnarray}
where $c_i$ ($i=1,2$)
are integration constants. From this relation, we infer that within our gravitational framework, the Barrow parameter is constrained to take integer values. According to the Barrow model, this condition would imply choosing either $\Delta = 0$ (corresponding to the Bekenstein-Hawking area law) or $\Delta = 1$. Although recent extensions of the Barrow model accommodating negative values of \( \Delta \) have been investigated~\cite{Bar9}, in this work we adhere strictly to the original formulation and focus on the case \( \Delta = 1 \). This choice allows us to maximize the cosmological impact of the extended entropy under investigation (for the choice of the remaining parameters and constants, we refer the reader to~\cite{fgtgrav}).

\begin{figure}[t]
\begin{center}
\includegraphics[width=8.5cm]{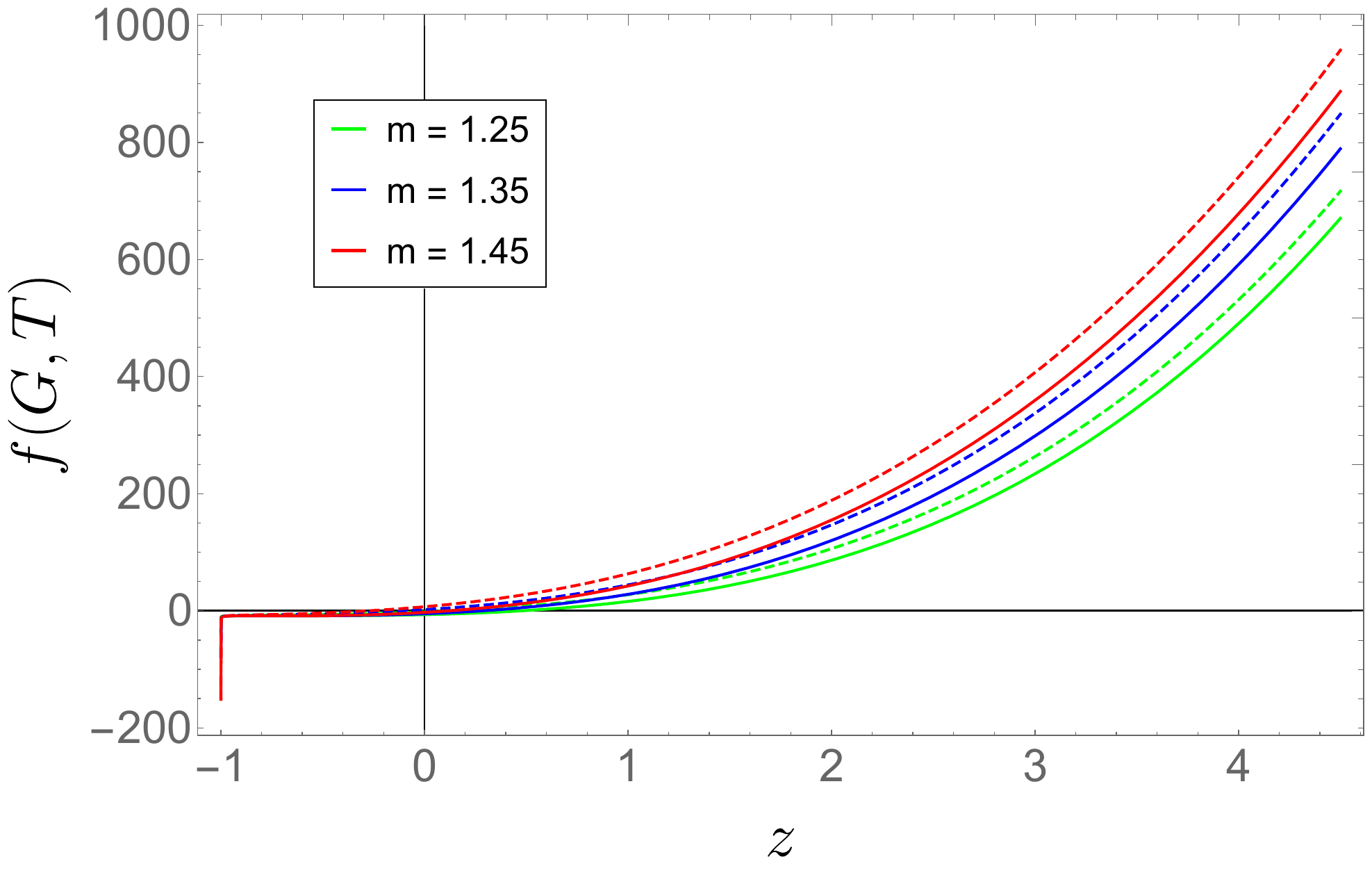}
\caption{\it{The evolution of $f(G,T)$ in Model I  for the parameter choice $c_1=0.25$, $c_2=-1.5$, $\rho_0=0.015$, $\eta_1=1.5$, $\eta_2=-3.2$  and $\Delta=1$ (solid lines). The corresponding curves for $\Delta=0$ are depicted with dashed lines. Different colors (online) represent distinct values of $m$, as indicated in the legend.}}
\label{fig1}
\end{center}
\end{figure}

As a final step, the reconstructed \( f(G, T) \) model is obtained by replacing Eqs.~\eqref{constra} and~\eqref{secf} in Eq.~\eqref{f1}. The resulting expression is given in Eq.~\eqref{Ap1} in the Appendix. The behavior of \( f(G,T) \) as a function of the redshift \( z = \dfrac{1 - a}{a} \) is plotted in Fig.~\ref{fig1} for different values of \( m > 1 \), in order to ensure consistency with a DE-dominated description of the Universe's expansion. 
We observe that the curves gradually decrease with decreasing redshift and approach zero at the present epoch \( z = 0 \), indicating a physically realistic model. Furthermore, it can be noted that higher values of the parameter \( m \) correspond to higher values of \( f(G,T) \), with the deviation from the standard \( \Delta = 0 \) curves (dashed lines) becoming increasingly pronounced at higher redshift. This behavior is consistent with the expectation that the most significant effects of this quantum gravity model would manifest during the early stages of the Universe's evolution.

\subsection{Model II}
The specific form of \( f(G, T) \) given in Eq.~\eqref{f2} corresponds to a non-conserved energy-momentum tensor scenario.  In fact, in this case, the coupling between curvature and matter (through the trace $T$) induces a non-trivial interaction. As a result, the field equations reduce to the form~\eqref{Heq}, with the corresponding energy density and pressure given by
\begin{eqnarray}
\label{newrhogt}
\rho_{GT}&=&\frac{3\eta T+F(G)}{2}-\frac{G}{2}F_G(G)+12H^3\dot G\hspace{0.2mm}F_{GG}(G)\,,\\[2mm]
p_{GT}&=&-\frac{\eta T+F(G)}{2}+\frac{G}{2}F_G(G)-4H\left[2\left(\dot H+H^2\right)\dot G+H\ddot G
\right]F_{GG}(G)-4H^2\dot G^2F_{GGG}(G)\,,
\label{pGT}
\end{eqnarray}
while the continuity equation~\eqref{cruc} becomes \cite{fgtgrav}
\be
\label{rho2}
\dot \rho+3H\rho\,=\,\left[-\frac{\eta}{2\left(1-\eta\right)}\right]\dot T\,\,\, \Longrightarrow\,\,\,
\rho\,=\,\rho_0\left(\tau-t\right)^{\frac{6m\left(1-\eta\right)}{\eta-2}}\,.
\ee

Following the same recipe as above, let us equate Eqs.~\eqref{rhode} and~\eqref{newrhogt}. By further using Eqs.~\eqref{at} and~\eqref{rho2},
after some algebra we get 
\begin{eqnarray}
\nonumber
&&G^2F_{GG}(G)+\frac{\left(m-1\right)G}{4}F_G(G)
-\frac{\left(m-1\right)}{4}F(G)
-\left[\frac{G}{24m^3\left(m-1\right)}\right]^{\frac{3m\left(\eta-1\right)}{2\left(\eta-2\right)}}\frac{3\eta\rho_0\left(m-1\right)}{4}\\[2mm]
&&+\left[\frac{C\left(m-1\right)}{2}\right]\left[-\frac{1}{24m^3}\left(\frac{G}{m-1}\right)^{\frac{1}{4}}\right]^{2-\Delta}\,=\,0\,,
\label{Gqu}
\end{eqnarray}
whose solution is provided by 
\begin{eqnarray}
&&\hspace{-5mm}F(G)\,=\,c_1G+c_2G^{\frac{1-m}{4}}+
\Bigg\{-m\hspace{0.2mm}C\hspace{0.2mm}
\left[-\frac{1}{12}\left(\frac{m}{m-1}\right)^{\frac{1}{4}}\right]^{-\left(2+\Delta\right)}
\bigg\{\left[m\eta^2\left(21m-17\right)+\eta\left(-45m^2+53m-8\right)\right.\\[2mm]
\nonumber
&&\hspace{-5mm}+\,2^{1-\frac{5}{4}(2+\Delta)}3^{-\frac{3}{4}(2+\Delta)}
\left(12m^2-19m+4\right)\big]+2\eta^2 864^{-\frac{2+\Delta}{4}}
\bigg\}\,G^{\frac{2-\Delta}{4}}+36\left(m-1\right)6^{-\frac{m(\eta-1)}{2(\eta-2)}}\left[\Delta-\left(m+1\right)\right]\left[\frac{1}{m^3\left(m-1\right)}\right]^{\frac{3m(\eta-1)}{2(\eta-2)}}\\[2mm]
\nonumber
&&\hspace{-5mm}\times\,48^{-\frac{m(\eta-1)}{\eta-2}}\,\eta\left(\eta-2\right)^2\rho_0\left(\frac{2+\Delta}{2}\right)G^{\frac{3m(\eta-1)}{2(\eta-2)}}
\Bigg\}\frac{1}{3\left(2+\Delta\right)\left[\Delta-\left(m+1\right)\right]\left[4+3m\left(\eta-1\right)-2\eta\right]\left[2-\eta+m\eta\left(7\eta-8\right)\right]}\,.
\label{GquSol}
\end{eqnarray}

\begin{figure}[t]
\begin{center}
\includegraphics[width=8.5cm]{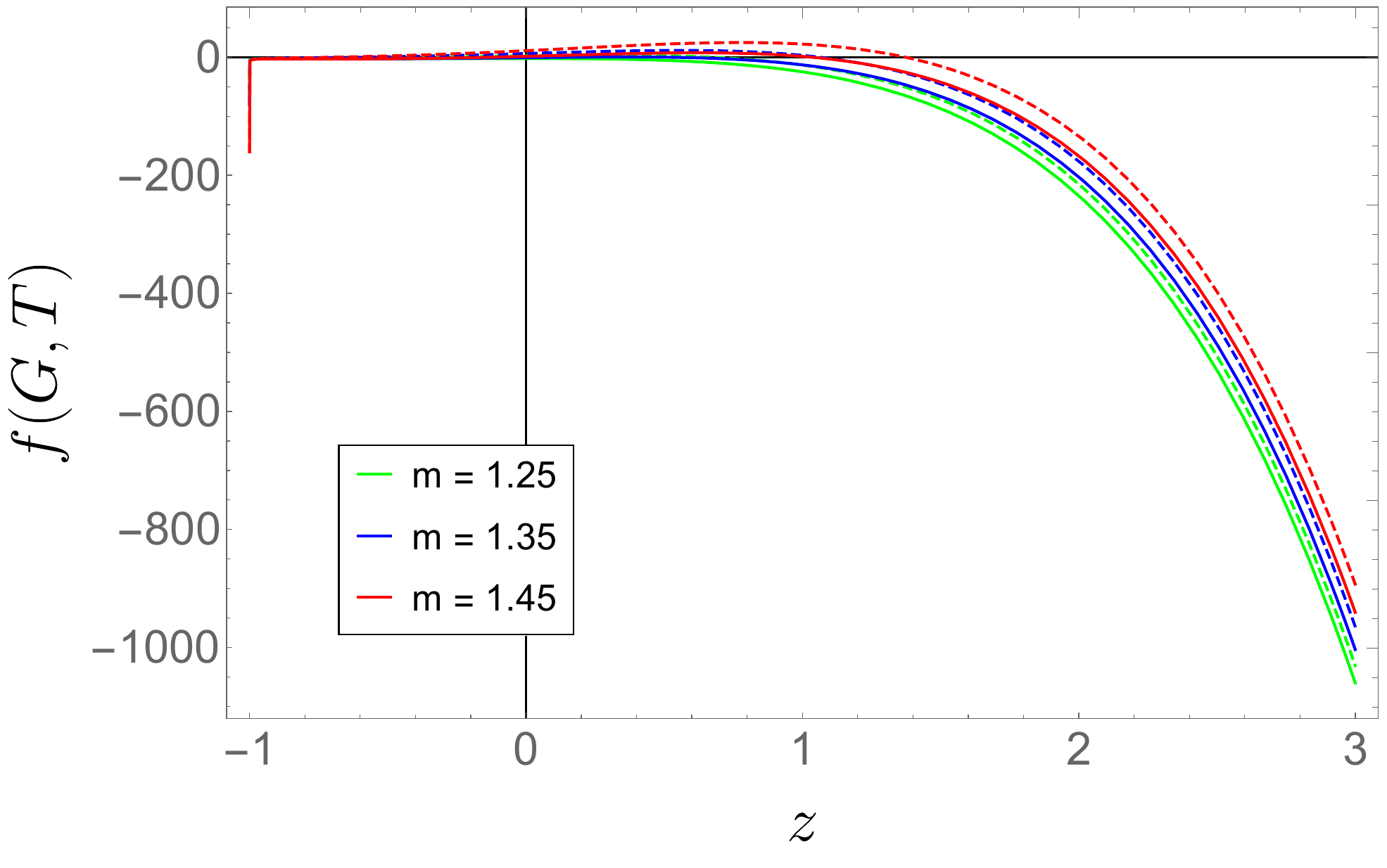}
\caption{\it{The evolution of $f(G,T)$ in Model II  for the parameter choice $c_1=0.25$, $c_2=-1.5$, $\rho_0=0.015$, $\eta=-3$  and $\Delta=1$ (solid lines). The corresponding curves for $\Delta=0$ are depicted with dashed lines. Different colors (online) represent distinct values of $m$, as indicated in the legend.}}
\label{fig2}
\end{center}
\end{figure}

The reconstructed BHDE $f(G,T)$ model is obtained by adding the term $\eta T$ to the above expression. The resulting function, which is presented in Eq.~\eqref{GquSol2} of the Appendix, is plotted against the redshift parameter in Fig.~\ref{fig2} for various values of $m$. The observed behavior, where the reconstructed $f(G,T)$ takes decreasing negative values with increasing redshift $z$, approaching zero as $z\rightarrow 0$, is physically meaningful within the framework of modified gravitational theories. 
Specifically, the fact that $f(G,T)$ is negative and its magnitude increases in the early Universe suggests an effective gravitational modification that acts analogously to a fluid with negative pressure or energy density. Such behavior could imply an effective repulsive gravitational interaction, potentially driving accelerated expansion phases (e.g., inflationary scenarios) in the early Universe. As the Universe evolves toward the present epoch ($z \rightarrow 0$), these gravitational modifications diminish, thereby recovering standard general relativity at late times, consistent with current observational constraints. Conversely, in scenarios 
where $f(G,T)$ is positive and grows with increasing redshift, as observed for Model I, 
gravitational modifications would manifest as enhanced gravitational attraction at early cosmic times. This latter framework could influence cosmic structure formation in a distinct manner, potentially giving rise to unique observational signatures in matter clustering and galaxy formation processes. 

From this perspective, a more comprehensive analysis would require examining the evolution of small density perturbations and the subsequent formation of large-scale structures in the early Universe. Indeed, the growth rate of these structures from initial density fluctuations constitutes one of the most rigorous observational probes for distinguishing among competing cosmological models~\cite{MatPe}. Specifically, it would be particularly valuable to explore whether, by suitably fine-tuning the model's free parameters, our extended gravitational framework can effectively contribute to resolving the well-known $\sigma_8$ tension between the amplitude of matter density fluctuations inferred from early-Universe observations and the comparatively lower values derived from late-time cosmological measurements~\cite{Abdalla:2022yfr}. A detailed investigation into this issue goes beyond the scope of the present work and will be addressed in a forthcoming study.

\section{Cosmological evolution of reconstructed  BHDE $f(G,T)$ models}
\label{Devot}
We now explore the cosmic evolution of the above 
reconstructed BHDE $f(G,T)$ models. Specifically, 
we analytically compute the EoS parameter, the deceleration parameter and the squared sound speed, and investigate the 
evolution trajectories of $\omega_{GT}-\omega'_{GT}$ phase plane. 

\subsection{Model I}
The equation-of-state (EoS) parameter is defined as the ratio
\be
\label{defom}
\omega_{GT}\,=\,\frac{p_{GT}}{\rho_{GT}},
\ee
which, within our $f(G,T)$ gravity-based reconstruction of the BHDE, coincides with $\omega_{GT}=p_{GT}/\rho_\Delta$.
By resorting to Eqs.~\eqref{rhogt},~\eqref{pGT},~\eqref{constra}
and~\eqref{secf}, we obtain the explicit expression of $\omega_{GT}$ as presented in Eq.~\eqref{om1} of the Appendix.

\begin{figure}[t]
\begin{center}
\includegraphics[width=8.5cm]{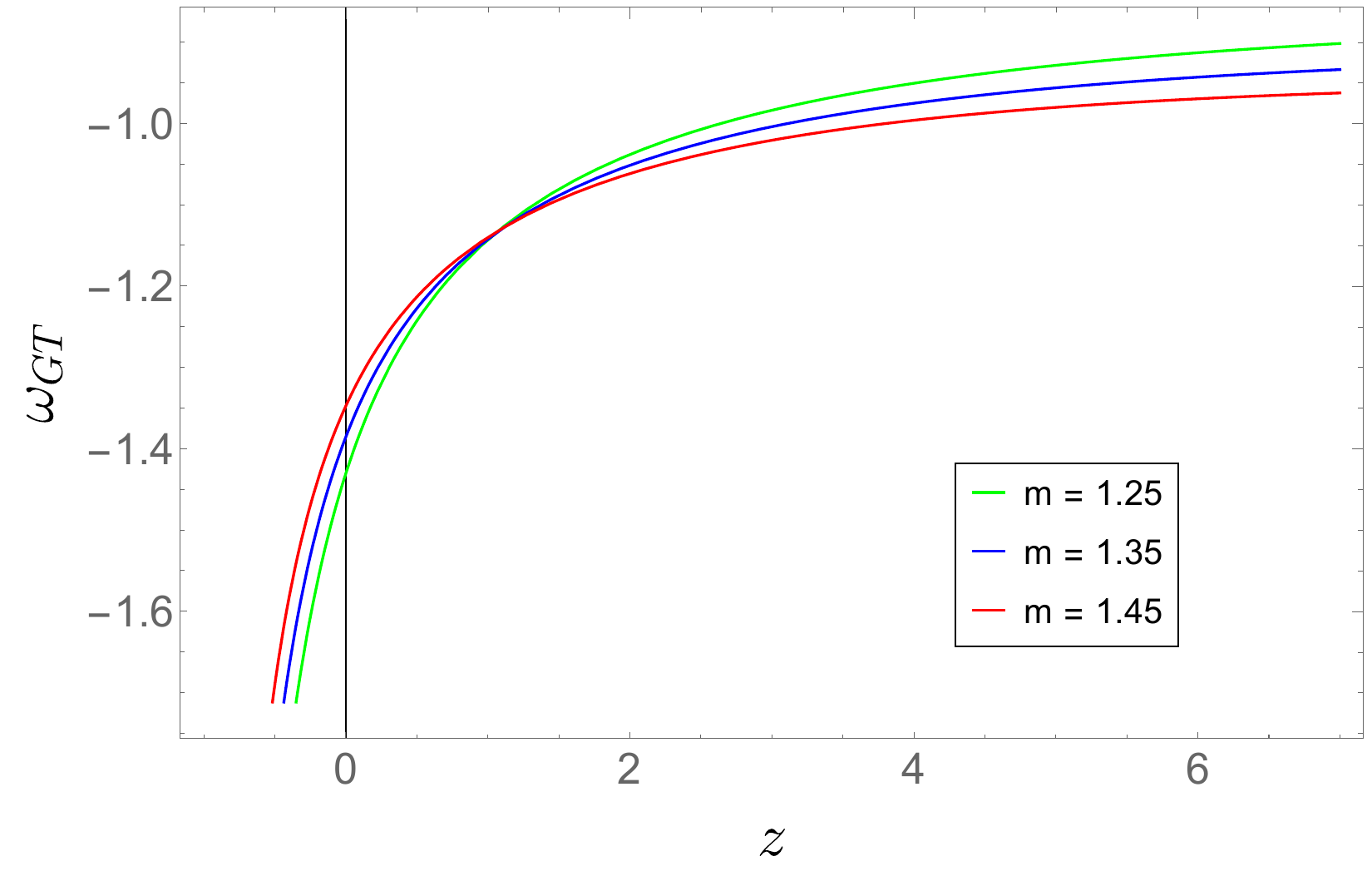}
\caption{\it{The evolution of $\omega_{GT}$ in Model I.  Different colors (online) represent distinct values of $m$, as indicated in the legend.}}
\label{fig3}
\end{center}
\end{figure}

The dynamics of $\omega_{GT}$ is plotted in Fig.~\ref{fig3}
for different values of $m$. We can see
that, in the present model, BHDE evolves from a quintessence-like 
phase ($-1<\omega_{GT}<-1/3$) at late time
to cosmological constant ($\omega_{GT}=-1$) and phantom regime ($\omega_{GT}<-1$) at present.
The ability of the dark energy component to exhibit a phantom-like equation of state at the current epoch represents a compelling feature of our model, as it offers a potential pathway for addressing the longstanding $H_0$ tension between early- and late-time cosmological observations~\cite{DiValentino:2020naf}. Specifically, phantom dark energy has been shown to enhance the rate of late-time cosmic acceleration, thereby increasing the value of $H_0$ inferred from Cosmic Microwave Background (CMB) data and alleviating the discrepancy with local determinations of the Hubble constant~\cite{Abdalla:2022yfr,LT}.

Furthermore, largely negative values of the
EoS parameter in the far future indicate that the Universe might end up
with a big-rip or remain in the same
current accelerating status. By comparison with~\cite{fgtgrav}, we
notice that such a behavior differs from the one exhibited
in $f(G,T)$ gravity-based reconstruction of Tsallis HDE, 
which is found to lie always in the 
phantom domain. On the other hand, a similar evolution 
is obtained within the framework of BHDE in Brans-Dicke Cosmology with a linear interaction~\cite{Bar7} and in Bianchi type-I BHDE in symmetric teleparallel gravity~\cite{TeleBar}. 
Quantitatively speaking, we observe that our model predicts $\omega_{{GT}_0}\in [-1.43,-1.34]$ for the considered values of $m$, which overlaps with $\omega_{DE_0}\in[-1.95,-1.03]$ suggested by multiple independent experimental probes (TT, TE, EE+lowE+lensing)~\cite{Planck}. 

To further enrich the cosmographic analysis, we now proceed to compute the deceleration parameter
\be
\label{qeq}
q\,=\,-\frac{\ddot a}{aH^2}\,=\,-1-\frac{\dot H}{H^2}\,.
\ee
This relation shows that positive values of $q$ correspond to a decelerated expansion of the Universe ($\ddot{a} < 0$), while negative values indicate an accelerated phase ($\ddot{a} > 0$). For the present model, the deceleration parameter is given by the expression in Eq.~\eqref{qeq2}.

\begin{figure}[t]
\begin{center}
\includegraphics[width=8.5cm]{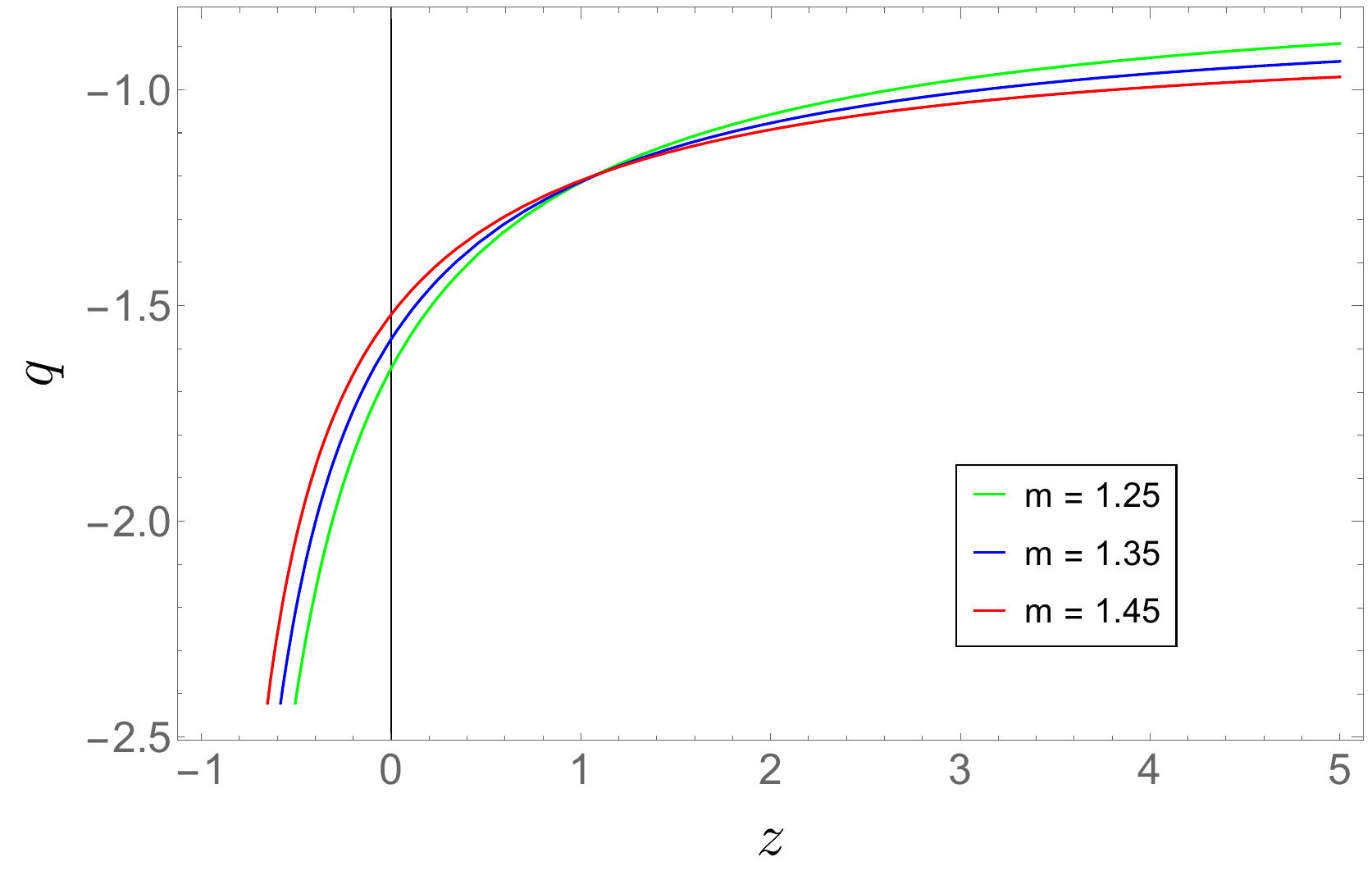}
\caption{\it{The evolution of $q$ in Model I.  Different colors (online) represent distinct values of $m$, as indicated in the legend.}}
\label{fig4}
\end{center}
\end{figure}

The evolution of the deceleration parameter $q$ is shown in Fig.~\ref{fig4}, demonstrating that the present model successfully captures the current accelerated expansion of the Universe. Notably, this represents a key advantage of the BHDE scenario with the original choice of the Hubble radius as the IR cutoff, in contrast to the standard HDE model, which fails to account for the observed late-time acceleration.
In particular, we find $q_0 \in [-1.64, -1.51]$ at the present epoch, a result that shows closer agreement with the estimate $q_0 = -1.08 \pm 0.29$ derived from supernovae observations~\cite{Camerana}, compared to the $\Lambda$CDM prediction of $q_0 \approx -0.55$~\cite{Planck}.
It should also be noted that the present model does not appear to reproduce the decelerated expansion phase ($q > 0$) of the early Universe. However, this limitation is expected to be naturally addressed by extending the present HDE model to include the radiation component in the Universe's energy budget. Indeed, incorporating radiation in BHDE description is essential for accurately capturing the early-time dynamics and ensuring a cosmic evolution that remains consistent with observations across all redshifts~\cite{SariBar1,Bar1}.

Let us now turn our attention to the squared sound speed, a key quantity in the analysis of cosmological models, as it determines how perturbations propagate within the cosmic fluid. Specifically, it describes the response of pressure to changes in energy density, thereby influencing the dynamics of structure formation and the overall stability of the model. In this context, the sign of $v_s^2$ becomes particularly relevant. A positive squared sound speed ($v_s^2 > 0$) ensures that pressure perturbations propagate as real sound waves, implying that small fluctuations remain under control and do not grow uncontrollably. This behavior is characteristic of a classically stable configuration. Conversely, a negative value of $v_s^2$ leads to imaginary sound speeds, which result in exponentially growing perturbations and potential instabilities~\cite{Pert1,Pert2}.

\begin{figure}[t]
\begin{center}
\includegraphics[width=8.8cm]{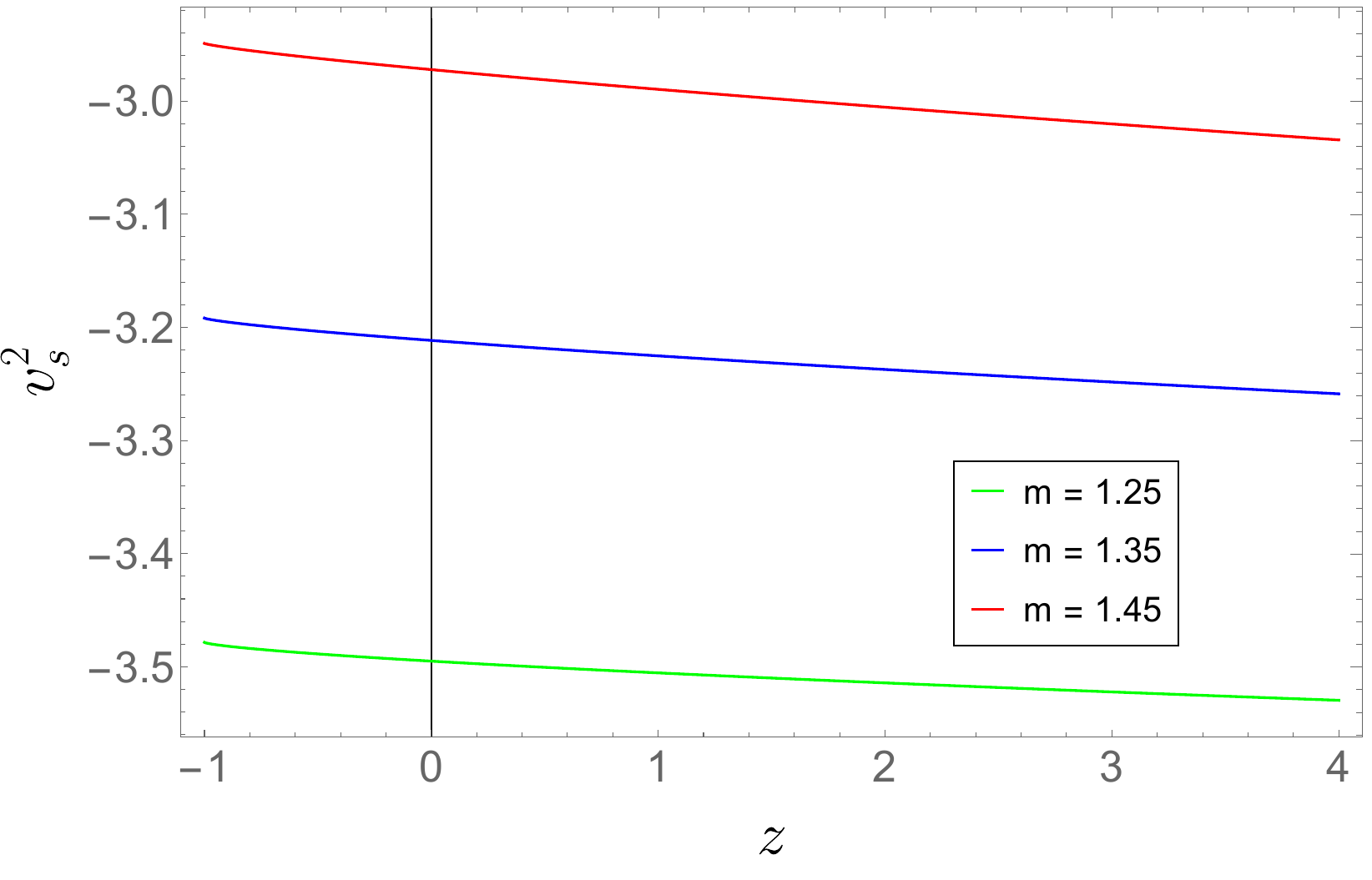}
\caption{\it{The evolution of $v_s^2$ in Model I.  Different colors (online) represent distinct values of $m$, as indicated in the legend.}}
\label{fig5}
\end{center}
\end{figure}

To assess this aspect within our framework, we consider the evolution of the squared speed of sound, defined as
\begin{equation} 
\label{vsq} 
v_s^2 = \frac{\dot{p}_{GT}}{\dot{\rho}_{GT}}\,.
\end{equation} 
By substituting Eqs.~\eqref{rhogt},~\eqref{pGT}, and~\eqref{cont1} into Eq.~\eqref{vsq}, and using the definition~\eqref{f1} of $f(G,T)$, we obtain the expression given in Eq.~\eqref{vsq1}. The evolution of $v_s^2$ is shown in Fig.~\ref{fig5} for different values of $m$, indicating that $v_s^2 < 0$ throughout the considered redshift range.

Although this result indicates the presence of classical instabilities, it is important to consider the broader context in which the model operates. First, it should be noted that the occurrence of negative $v_s^2$ over a given redshift range may reflect a transient feature of the model rather than a fundamental pathology. If the background evolution remains well-behaved and the perturbations do not generate significant effects, such a feature can be tolerated within acceptable theoretical margins~\cite{Pert1,Pert2}.
Most importantly, the analysis of $v_s^2$ should be complemented by a full perturbative treatment that includes both scalar and metric perturbations, in order to accurately assess the severity of the instability. In fact, in some cases, additional dynamical mechanisms - such as modifications to the effective sound speed, entropy perturbations or non-adiabatic effects - may act to stabilize the evolution at the perturbative level, even when $v_s^2 < 0$ arises within the adiabatic approximation.
Lastly, we note that a similar feature also arises in other dark energy reconstruction models based on extended entropies - for instance, in the case of Tsallis holographic dark energy~\cite{fgtgrav}, where the squared sound speed remains negative over a comparable redshift range. This indicates that classical instabilities may be a common aspect of related theoretical frameworks, further motivating a deeper investigation into their underlying dynamics and possible stabilizing mechanisms.

Let us finally analyze the trajectories in the \(\omega_{GT} - \omega_{GT}'\) plane, where the prime denotes the derivative with respect to \(\log a\), i.e.
\begin{equation}
\omega_{GT}'\,\equiv\,\frac{d\omega_{GT}}{d \log a}\,=\, -\frac{d\omega}{dz}\left(1+z\right)\,.
\end{equation}
This phase-space representation provides a powerful diagnostic tool for understanding the dynamical behavior of dark energy models through the evolution of their effective equation of state. Originally introduced in~\cite{Cald} in the context of quintessence, the \(\omega_{GT} - \omega_{GT}'\) plane enables the classification of models based on whether they exhibit thawing behavior (\(\omega_{GT}<0\), \(\omega_{GT}'>0\)) or freezing behavior (\(\omega_{GT}<0\), \(\omega_{GT}'<0\)). This framework has since been extended to a broad class of dynamical dark energy models, including those with non-canonical kinetic terms and entropic corrections~\cite{Sche,Chiba,Guo,Sharif}.

\begin{figure}[t]
\begin{center}
\includegraphics[width=8cm]{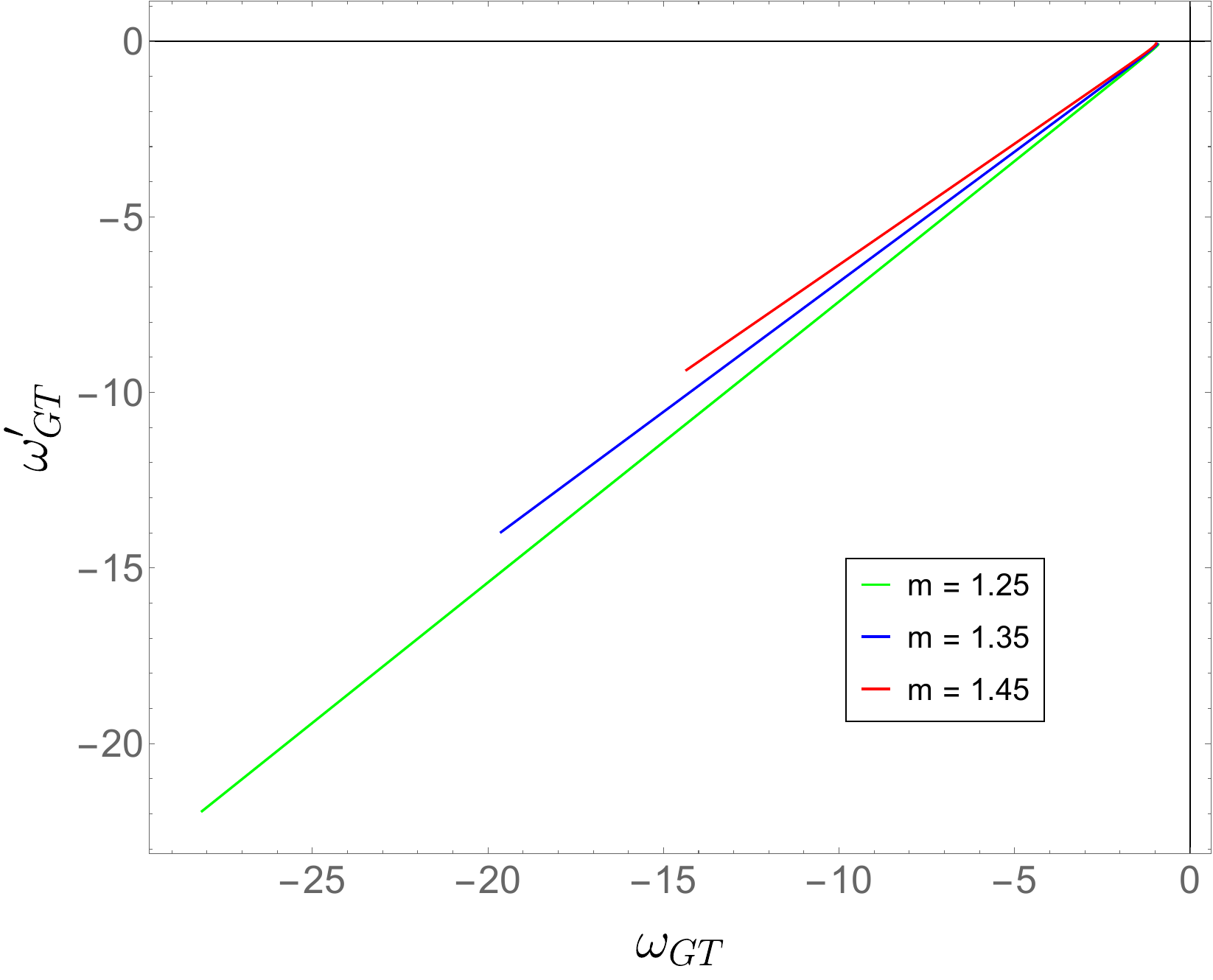}
\caption{\it{The evolution of the $\omega_{GT}-\omega'_{GT}$ trajectories in Model I.  Different colors (online) represent distinct values of $m$, as indicated in the legend.}}
\label{fig6}
\end{center}
\end{figure}

The cosmic trajectories in the $\omega_{GT}$-$\omega'_{GT}$ plane for the specific values of $m$ considered above are displayed in the parametric plot of Fig.~\ref{fig6}. The figure reveals that the present model evolves within the freezing region. This behavior is particularly relevant, as the freezing regime is typically associated with a more sustained and stronger acceleration of the Universe's expansion compared to the thawing region. Moreover, this outcome aligns well with previous analyses in related frameworks, such as the Tsallis holographic dark energy model~\cite{fgtgrav} and the Bianchi type-I BHDE scenario in symmetric teleparallel gravity~\cite{TeleBar}.

\subsection{Model II}
We now investigate the cosmic evolution of the reconstructed BHDE \( f(G,T) \) model based on Eq.~\eqref{f2}. Following the same procedure as in the previous case, we use Eq.~\eqref{defom} along with the definitions~\eqref{newrhogt} and~\eqref{pGT} to derive the equation-of-state parameter, whose expression is given in Eq.~\eqref{EoS2} of the Appendix. 

\begin{figure}[t]
\begin{center}
\includegraphics[width=8.5cm]{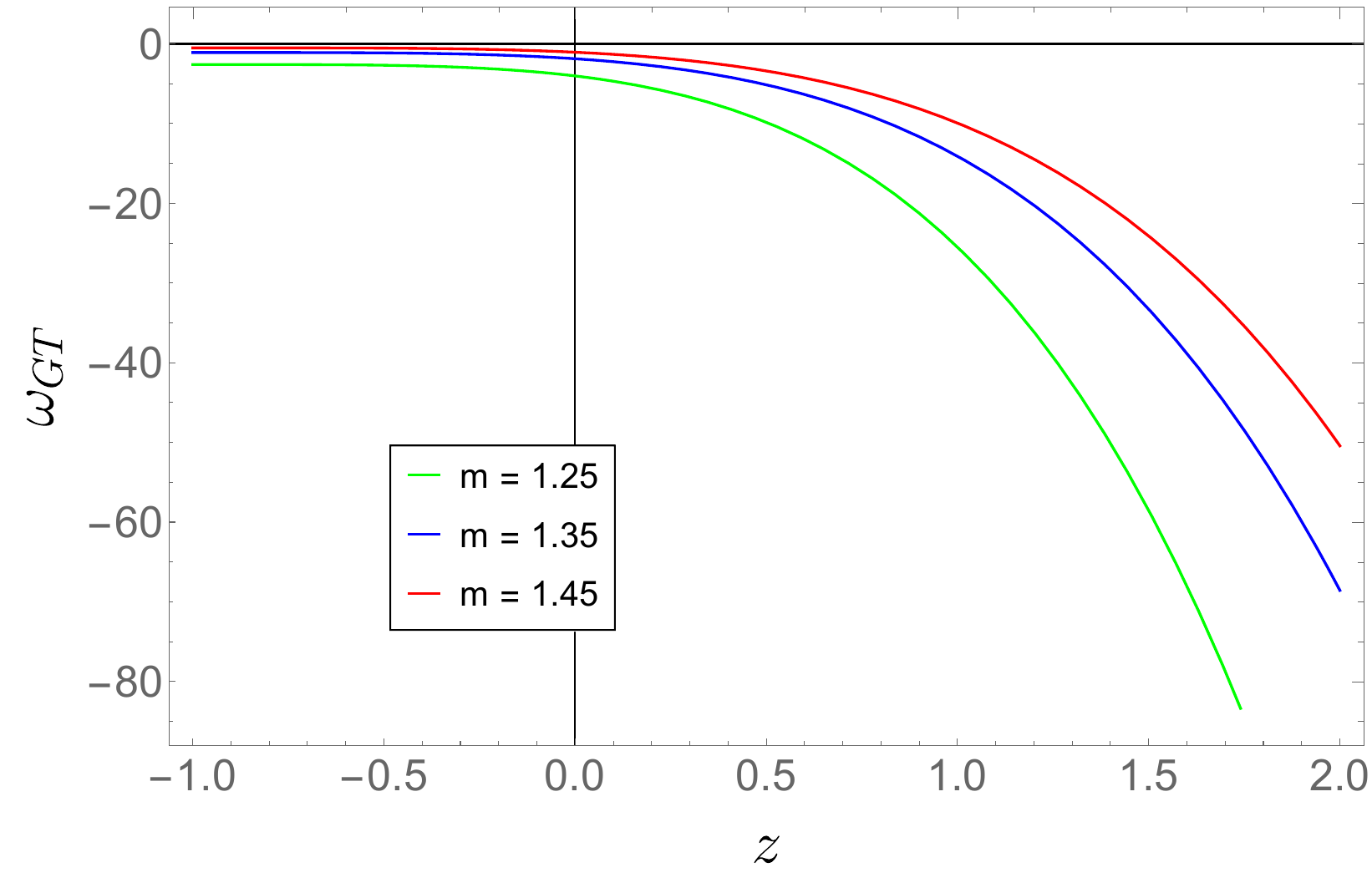}
\caption{\it{The evolution of $\omega_{GT}$ in Model II.  Different colors (online) represent distinct values of $m$, as indicated in the legend.}}
\label{fig7}
\end{center}
\end{figure}

The evolution of \( \omega_{GT} \) is shown in Fig.~\ref{fig7} for different values of \( m \). Unlike Model~I, the reconstructed BHDE model lies in the phantom regime during the early Universe, evolving either toward a cosmological constant-like behavior for \( m = 1.35 \), or toward a quintessence-like regime for \( m = 1.45 \) as the redshift \( z \) decreases. For these values of \( m \), the present-day equation-of-state parameter lies within the range \( \omega_{GT_0} \in [-1.86, -1.05] \), which remains in good agreement with current observational data~\cite{Planck}.

\begin{figure}[t]
\begin{center}
\includegraphics[width=8.5cm]{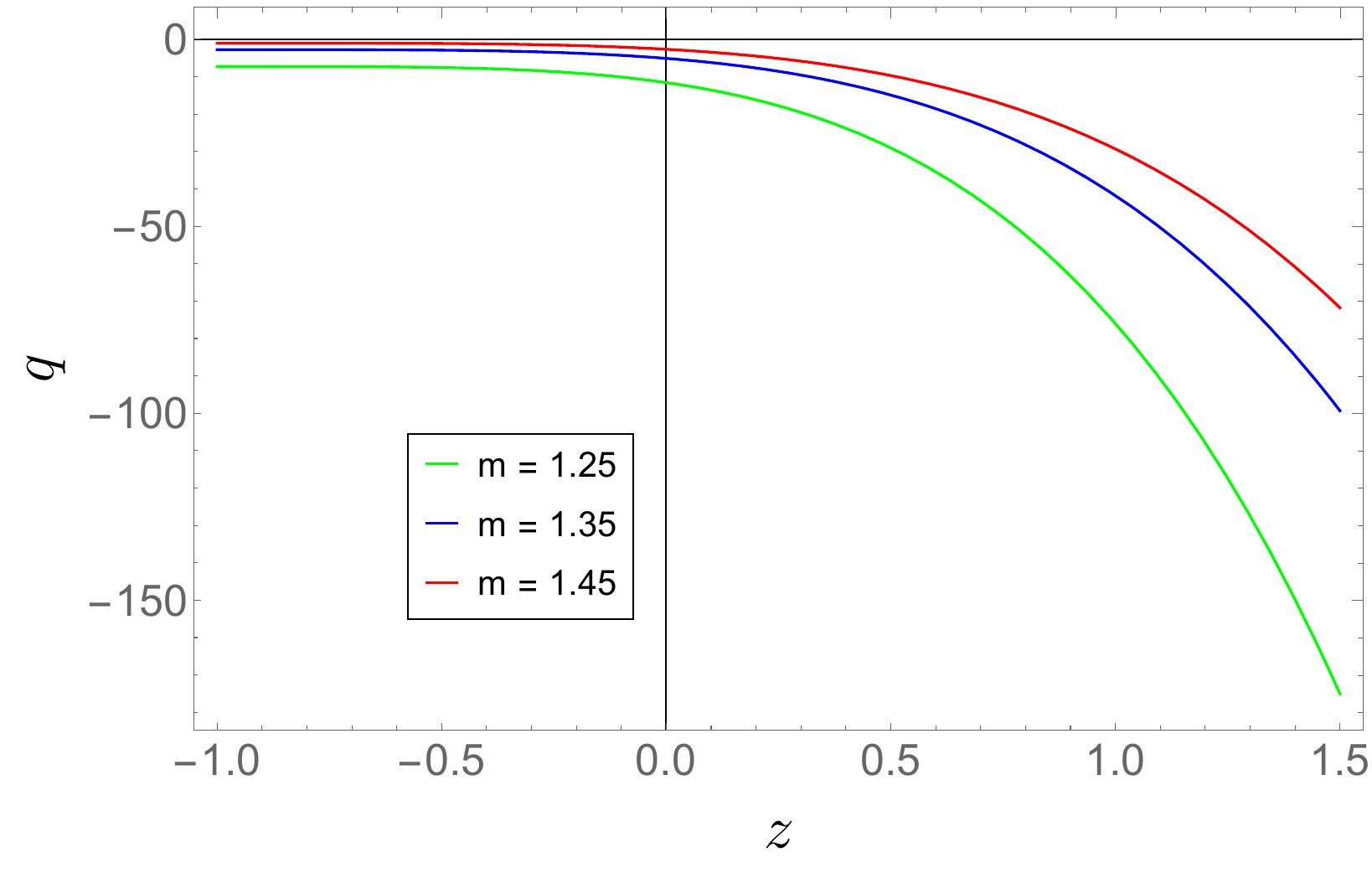}
\caption{\it{The evolution of $q$ in Model II.  Different colors (online) represent distinct values of $m$, as indicated in the legend.}}
\label{fig8}
\end{center}
\end{figure}

In turn, the corresponding expression for the deceleration parameter is presented in Eq.~\eqref{dec2}, with its evolution depicted in Fig.~\ref{fig8}. As in Model~I, we find that \( q < 0 \), indicating an accelerating phase of cosmic expansion. However, for the considered values of the model parameters, the present-day deceleration parameter is found to obey \( q_0 \lesssim -1.81 \), which appears to deviate from current observational estimates~\cite{Planck, Camerana}. This deviation may indicate limitations in the parameter selection or foundational assumptions of Model~II, 
suggesting that further refinement or alternative formulations may be necessary to achieve better 
phenomenological consistency. In this context, Model~I demonstrates greater compatibility with current observational constraints.

\begin{figure}[t]
\begin{center}
\includegraphics[width=9cm]{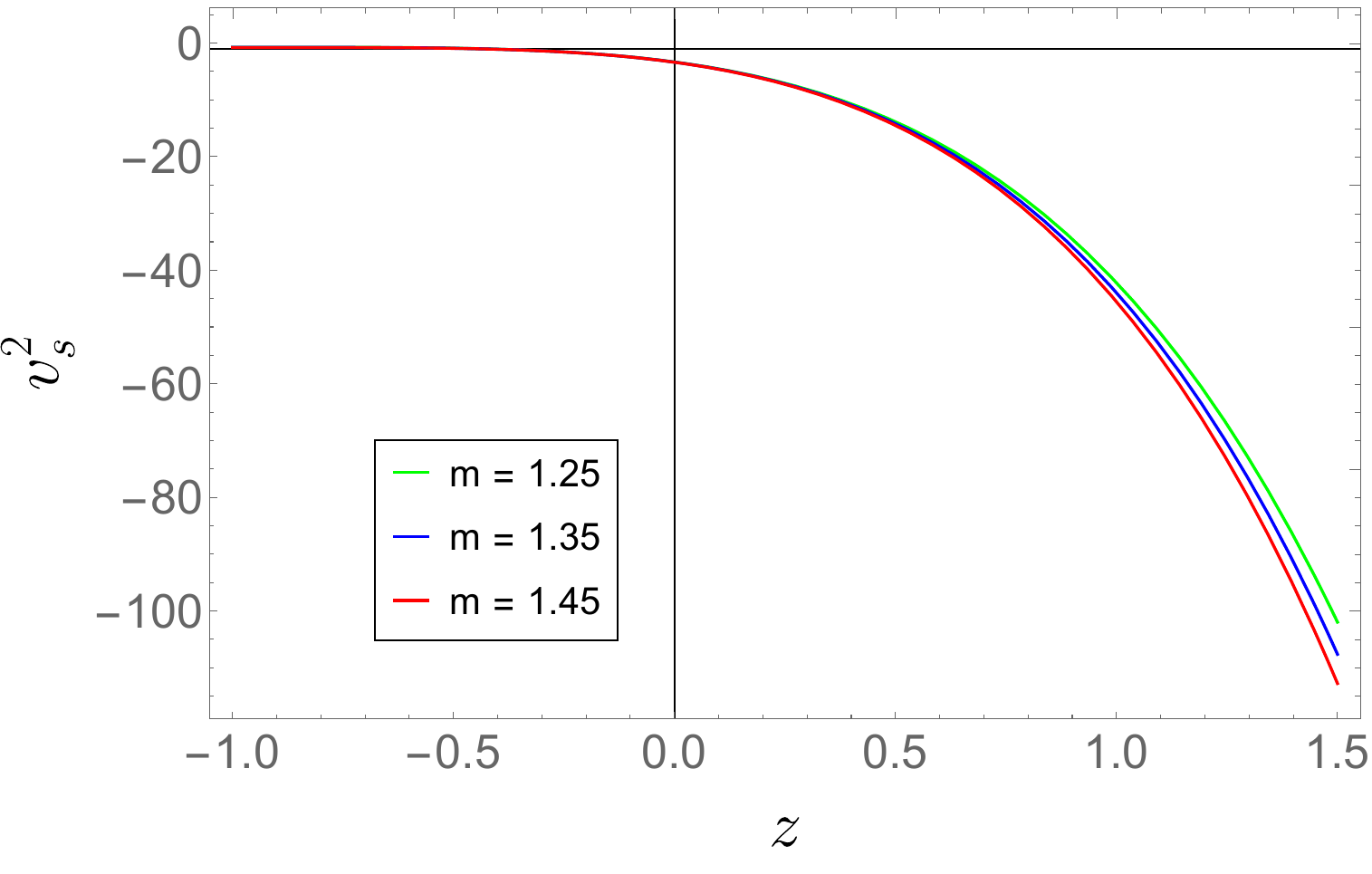}
\caption{\it{The evolution of $v_s^2$ in Model II.  Different colors (online) represent distinct values of $m$, as indicated in the legend.}}
\label{fig9}
\end{center}
\end{figure}

The stability analysis of Model II is presented in Fig.~\ref{fig9}, based on the analytic expression of $v_s^2$ provided in Eq.~\eqref{vs2}. 
Interestingly, the results indicate that the reconstructed BHDE model evolves from an unstable configuration ($v_s^2<0$) in the early Universe to a stable state ($v_s^2>0$) in the far future. 

\begin{figure}[t]
\begin{center}
\includegraphics[width=8.7cm]{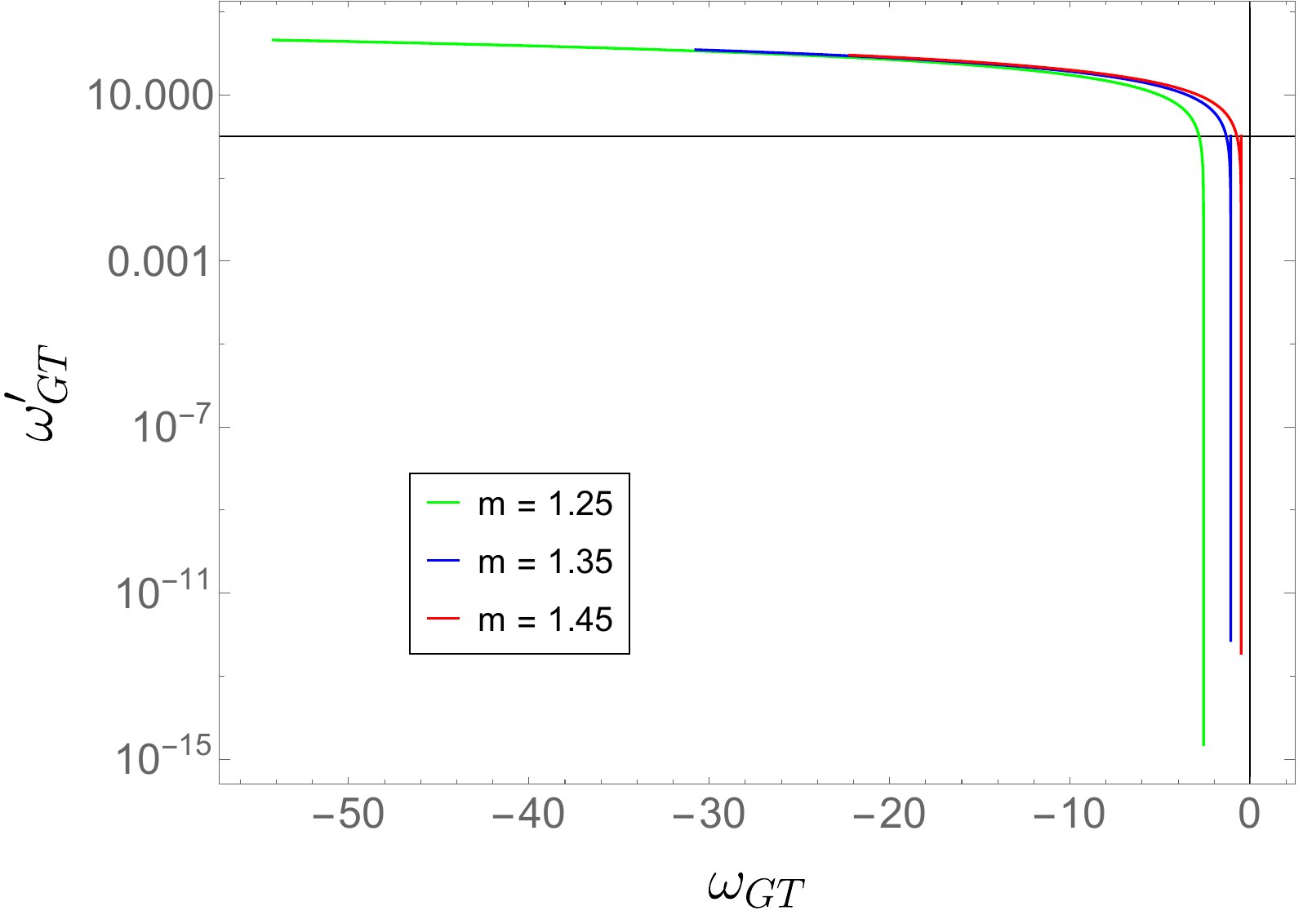}
\caption{\it{The evolution of $\omega_{GT}-\omega'_{GT}$ trajectories in Model II.  Different colors (online) represent distinct values of $m$, as indicated in the legend.}}
\label{oopr}
\end{center}
\end{figure}

Let us finally turn our attention to the cosmic trajectories in the \( \omega_{GT} \)-\( \omega'_{GT} \) phase space for the specific values of the parameter \( m \) considered in this analysis. These trajectories are depicted in Fig.~\ref{oopr}, which indicates that the present model evolves within the thawing region. In this regard, we remark that  thawing models are typically associated with scalar field dynamics in which the equation-of-state parameter departs from the cosmological constant value \( \omega = -1 \) at late times, eventually entering the phantom regime. This is in line with the dynamics displayed in Fig.~\ref{fig7}. Such evolution lends further support to the phantom dark energy scenario, which has been proposed as a viable explanation for the observed late-time acceleration of the Universe.

\section{Observational Constraints}
\label{obc}

As discussed above, Model I is favored over Model II based on its better agreement with observational data, particularly in terms of the present-day value of the deceleration parameter. To further assess the observational viability of this framework and place constraints on the parameters \( \Delta \), \( m \), and \( H_0 \), we analyze the evolution of the Hubble parameter \( H(z) \), as obtained from Eqs.~\eqref{Heq} and~\eqref{rho}, by fixing the values of \( \eta_1 \), \( \eta_2 \) and \( \rho_0 \) as specified in Sec.~\ref{reconef}. The resulting theoretical prediction is then compared with a dataset comprising 57 measurements of the Hubble parameter in the redshift range \( 0.07 \le z \le 2.36 \). These data points are drawn from a variety of sources, including 31 determinations based on the Differential Age (DA) method and 26 obtained via Baryon Acoustic Oscillations (BAO) and other independent observational techniques~\cite{TeleBar}.

As in~\cite{TeleBar}, we use the statistical \( R^2 \)-test to determine the best-fit values of the model parameters:
\begin{equation}
R^2 = 1 - \frac{\sum_{i=1}^{57} \left[(H_i)_{\text{ob}} - (H_i)_{\text{th}}\right]^2}{\sum_{i=1}^{57} \left[(H_i)_{\text{ob}} - (H_i)_{\text{mean}}\right]^2} \,,
\label{Rq}
\end{equation}
where \( (H_i)_{\text{ob}} \) and \( (H_i)_{\text{th}} \) denote the observed and theoretical values of the Hubble parameter, respectively. The quantity \( R^2 \) represents the coefficient of determination and quantifies the fraction of the variance in the observational data explained by the model; values close to 1 indicate a good fit.

By minimizing the sum of squared residuals in the numerator of Eq.~\eqref{Rq}, we obtain the best-fit parameters \( m = 1.8 \), in agreement with the onset of a cosmic acceleration era (as discussed below Eq.~\eqref{at}) and \( \Delta \sim \mathcal{O}(10^{-1}) \). 
Although the best-fit value of \( \Delta \) is not an integer - as formally required by the present theoretical formulation (see below Eq.~\eqref{f1Gne}) - this result remains fully compatible with the physical expectations of the model. Indeed, the restriction to integer values arises from the need to preserve consistency within the analytic structure imposed by the specific power-law ansatz~\eqref{at}, rather than from any fundamental limitation of the underlying theory. Moreover, the estimated value lies well within the original Barrow range \( [0,1] \), consistent with the scale of quantum-gravitational corrections, and may be interpreted as an effective parameter capturing the dominant physical effects within this framework. 

Interestingly, our estimate is consistent with the result \( \Delta = 0.094^{+0.093}_{-0.101} \), obtained in~\cite{Anagnostopoulos:2020ctz} from a phenomenological analysis of holographic dark energy based on observational data from the SNIa Pantheon sample and direct Hubble parameter measurements via the cosmic chronometers (CC) method. However, it turns out to be less stringent than the bounds \( 0.005 \leq \Delta \leq 0.008 \)~\cite{Bar6} and \( \Delta \lesssim 1.4 \times 10^{-4} \)~\cite{Barrow:2020kug}, which were derived from baryogenesis and Big Bang Nucleosynthesis constraints, respectively. It is important to emphasize, nonetheless, that these tighter constraints arise from studies of deformed cosmological models, rather than from direct applications of the holographic dark energy framework.

\begin{figure}[t]
\begin{center}
\includegraphics[width=8.5cm]{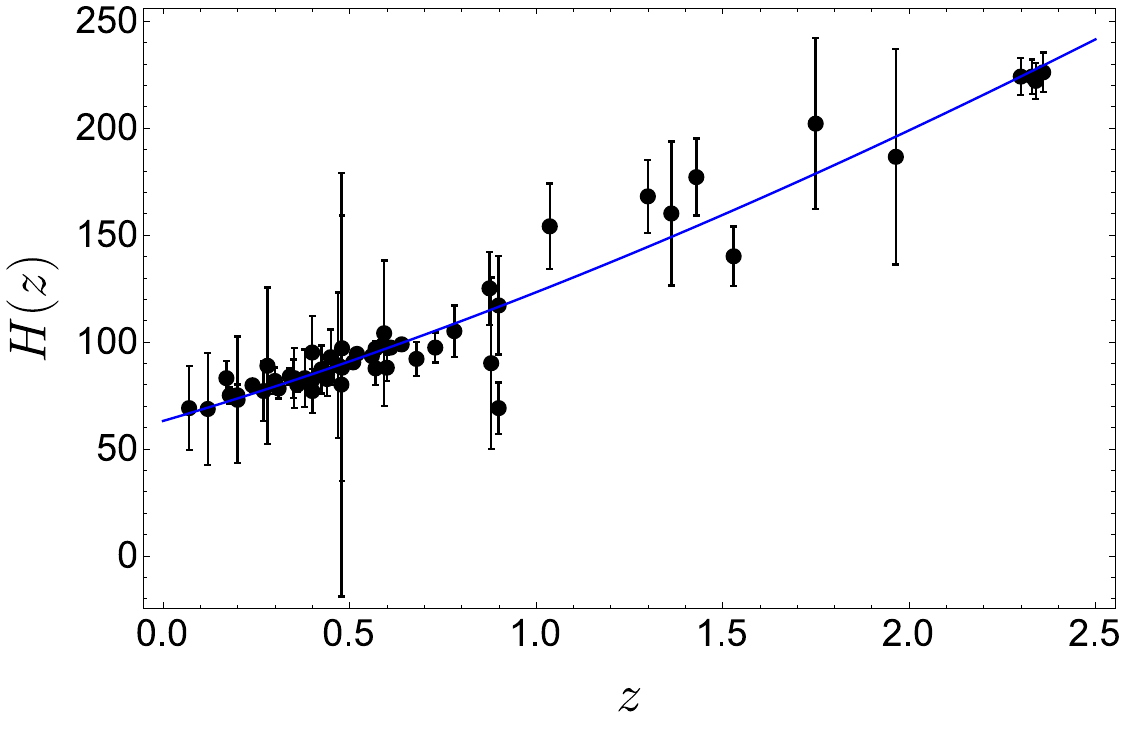}
\caption{\it{Best fit curve of Hubble's parameter $H$ versus redshift $z$. Dots represent observed values, while the blue curve is the theoretical fit.}}
\label{PlotH}
\end{center}
\end{figure}

From the comparison with observational data presented in Fig.~\eqref{PlotH}, we can also extract a present-day value of the Hubble parameter of \( H_0 = 65.1\,\mathrm{km\,s^{-1}\,Mpc^{-1}} \). This value is more closely aligned with the estimate reported by the Planck Collaboration, \( H_0 = \left(67.27 \pm 0.60\right)\,\mathrm{km\,s^{-1}\,Mpc^{-1}} \)~\cite{Planck}, derived from early-Universe Cosmic Microwave Background measurements, than with the higher local determination \( H_0 = \left(73.04 \pm 1.04\right)\,\mathrm{km\,s^{-1}\,Mpc^{-1}} \), obtained from distance-ladder measurements based on Cepheids and Type Ia supernovae~\cite{Riess:2021jrx}.

 \begin{center}
\begin{table}[t]
    \begin{tabular}{|c|c|c| c|}
    \hline
 &\,\,Model I \,\, &\,\, Model II  \,\,&\,\, Experimental constraint\,\, \\
 \hline
 \hline
\,\,$\omega_{GT_0}$ \,\, &\,\, $[-1.43,-1.34]$ &\,\,  $[-1.86,-1.05]$ \,\,&\,\, $[-1.95,-1.03]$\, \cite{Planck} \,\, \\
           \hline
     \,\,      $q_0$ \,\, &\,\, $[-1.64,-1.51]$ &\,\,  $[-10.95,-1.81]$ \,\,&\,\, $[-1.37,-0.79]$\,~\cite{Camerana} \,\,\\
           \hline
                    \,\,    $\omega_{GT}-\omega'_{GT}$ \,\, &\,\, freezing   &\,\,  thawing  \,\,&\,\, - \,\,\\
            \hline
        \,\,    $H_0$ \,\, &\,\, $65.1\,\mathrm{km\,s^{-1}\,Mpc^{-1}}$  &\,\,  $-$  \,\,&\,\, $\left(67.27\pm0.60\right)\mathrm{km\,s^{-1}\,Mpc^{-1}}$~\cite{Planck}\,\,\\
            \hline
    \end{tabular}
  \caption{\it{Theoretical and observational values of some cosmological parameters in the current epoch.}}
  \label{TabII}
\end{table}
\end{center}

\section{Conclusions and outlook}
\label{Conc}
We have developed a reconstruction of Barrow Holographic Dark Energy (BHDE) within the framework of \( f(G,T) \) gravity, using the Hubble horizon as IR cutoff. This study is motivated by the fact that BHDE offers a nontrivial approach to explore quantum gravity effects on the cosmological evolution of the Universe, despite the absence of a corresponding Lagrangian formulation. In this regard, \( f(G,T) \) gravity provides an ideal setting for implementing the reconstruction of BHDE, as it has proven to be a versatile framework for addressing several fundamental issues inherent in General Relativity, particularly those connected to the possible quantum nature of gravity~\cite{fgt1,fgt2,fgt3,fgt4}.

As the background geometry for our analysis, we considered a spatially flat Friedmann-Robertson-Walker (FRW) spacetime, with a dust fluid configuration, under both conserved and non-conserved energy-momentum tensor scenarios. To trace the evolutionary dynamics of the Universe in this context, we adopted a power-law ansatz for the scale factor and investigated the behavior of key cosmological parameters, including the equation-of-state (EoS) parameter, the deceleration parameter and the squared sound speed. Additionally, we analyzed the trajectories in the \( \omega_{GT} \)-\( \omega'_{GT} \) phase plane. For the conserved case (Model I), we found that the BHDE model exhibits a transition from the quintessence regime (\( -1 < \omega_{GT} < -1/3 \)) to the phantom regime (\( \omega_{GT} < -1 \)) as it approaches the present epoch. Moreover, the model consistently supports a late-time accelerated expansion phase. This behavior represents a notable advantage of the present model over the standard Holographic Dark Energy framework, which is unable to account for the current phase of cosmic acceleration. In addition, the \( \omega_{GT} \)-\( \omega'_{GT} \) trajectories fall within the freezing region, further supporting the viability and dynamical consistency of the BHDE model within the \( f(G,T) \) gravity framework.

On the other hand, in the non-conserved case (Model II), the BHDE model exhibits phantom-like behavior in the early Universe, subsequently evolving toward either a cosmological constant-like state or a quintessence-like regime at late times. Furthermore, the $\omega_{GT}-\omega'_{GT}$ trajectory indicates a
thawing behavior.

Finally, we have examined the phenomenological consistency of the obtained results, showing that Model I is favored over Model II by observational data, particularly with respect to the present value of the deceleration parameter. Accordingly, we have employed Model I to constrain the free parameters of the framework by analyzing the evolution of \( H(z) \). This was achieved by fitting observational data comprising 57 measurements of the Hubble parameter over the redshift range \( 0.07 \le z \le 2.36 \), using the statistical \( R^2 \)-test as the fitting criterion. 
The best-fit analysis yielded \( m = 1.85 \) and \( \Delta \sim \mathcal{O}(10^{-1}) \). From the corresponding best-fit curve, we also inferred a Hubble constant of \( H_0 = 65.1\,\mathrm{km\,s^{-1}\,Mpc^{-1}} \) within the framework of the present model.

Several aspects remain to be explored. First, since our model incorporates quantum gravitational corrections into cosmology through Barrow entropy, it is essential to examine the results in relation to the predictions of more fundamental candidate theories of quantum gravity.
Furthermore, in line with the approaches discussed in~\cite{Mamon:2020spa,Chakraborty:2021uzp}, it would be worthwhile to investigate the thermodynamic implications of our model and to assess its thermal stability. This analysis is crucial for determining whether the reconstructed BHDE within \( f(G,T) \) gravity can serve as a viable candidate for explaining the still unknown nature of dark energy. As an additional perspective, it would also be of interest to extend the present framework to alternative holographic dark energy models based on generalized entropies, in order to explore whether they exhibit common features or distinct behaviors within modified gravity scenarios.

From an observational standpoint, a natural extension of this work involves exploring the influence of our modified gravity model on the evolution of matter perturbations. 
In this context, a key objective of our forthcoming work is to investigate whether the present framework can offer insights into the long-standing \( \sigma_8 \) tension-namely, the discrepancy between the amplitude of matter density fluctuations inferred from early-Universe observations and the lower values suggested by late-time cosmological data~\cite{Abdalla:2022yfr}.
Given the modifications introduced by our extended gravitational model to the Hubble expansion history, it is reasonable to anticipate notable effects on the evolution of matter inhomogeneities. This expectation is supported by analogous findings in various alternative gravity theories with extended entropies ~\cite{Sheykhi:2022gzb,Basilakos:2023kvk,Luciano:2025ezl}. Additionally, observational constraints from large-scale structure data - such as galaxy clustering and weak gravitational lensing surveys - can be employed to quantitatively assess potential deviations from the predictions of the standard cosmological model. These probes provide complementary information on the growth of cosmic structures and can play a crucial role in testing the viability of extended gravity.

Finally, an intriguing direction involves exploring possible connections between holographic dark energy models formulated within extended gravity frameworks and dark energy phenomena emerging from exotic sectors of particle physics. In particular, the role of neutrinos has garnered significant attention~\cite{N1,N2,N3,N4}, especially in scenarios where their masses vary dynamically due to interactions with a scalar field driving cosmic acceleration. These models, commonly referred to as mass-varying neutrino scenarios, posit that the mechanism responsible for neutrino mass generation is deeply intertwined with the dark energy sector. This coupling can lead to nontrivial modifications of the background cosmological evolution, potentially offering novel solutions to outstanding problems such as the coincidence problem or late-time deviations from \(\Lambda\)CDM. Establishing a theoretical correspondence between such particle physics-inspired mechanisms and holographic dark energy within the context of modified gravity could pave the way for a unified phenomenological framework. This approach may not only deepen our understanding of dark energy but also provide testable predictions for future observational surveys. Ongoing work is directed toward developing these ideas further, and the results will be presented in forthcoming studies.

\appendix

\section{} 
\label{AppMath}
This appendix provides mathematical details and explicit derivations related to the reconstruction of BHDE within the framework of $f(G,T)$ gravity. 

\subsection{Reconstructing $f(G,T)$ gravity}

In this section, we aim to reconstruct the functional form of the gravitational action within the framework of $f(G,T)$ gravity. For Model~I, described by Eq.~\eqref{f1}, the full expression for $f(G,T)$ can be obtained by combining the constraints from Eqs.~\eqref{constra} and~\eqref{secf}, resulting in
\begin{eqnarray}
    \label{Ap1}
    \nonumber
    &&f(G,T)=c_1G+c_2G^{\frac{1-m}{4}}+\frac{1}{12\left(m+3\right)}
\left\{54^{-\frac{(2+\Delta)}{4}}C\hspace{0.2mm}m\hspace{0.4mm}\left[-\frac{m}{6\left[m^3\left(m-1\right)\right]^{\frac{1}{4}}}\right]^{-(2+\Delta)}
\left[\frac{4\left(3+m\right)}{\left(1+m-\Delta\right)\left(2+\Delta\right)}
\right]G^{\frac{2-\Delta}{4}}\right.\\[2mm]
\nonumber
&&+\,24\left(\frac{2}{3}\right)^{\frac{m}{8}}\hspace{-1mm}\eta_1{\rho_0^{\frac{1}{2}}}\left(m-1\right)\left[m^3\left(m-1\right)\right]^{-\frac{3m}{8}}\left[2^{-m}\left(\frac{8}{3m-8}\right)-76^{-\frac{m}{4}}\left(\frac{8}{5m-2}\right)
\right]G^{\frac{3m}{8}}-12\eta_2\left(m-1\right)\left(\frac{m-5}{m-1}\right)
\Bigg\}\nonumber\\[2mm]
&&+\,\eta_1 T^{1/2}+\eta_2\,.
\end{eqnarray}

On the other hand, for Model II presented in Eq.~\eqref{f2}, the reconstructed BHDE $f(G,T)$ is obtained by adding the term $\eta T$ to Eq.~\eqref{GquSol}, yielding
\begin{eqnarray}
&&\hspace{-5mm}f(G,T)\,=\,c_1G+c_2G^{\frac{1-m}{4}}+
\Bigg\{-m\hspace{0.2mm}C\hspace{0.2mm}
\left[-\frac{1}{12}\left(\frac{m}{m-1}\right)^{\frac{1}{4}}\right]^{-\left(2+\Delta\right)}
\bigg\{\left[m\eta^2\left(21m-17\right)+\eta\left(-45m^2+53m-8\right)\right.\\[2mm]
\nonumber
&&\hspace{-5mm}+\,2^{1-\frac{5}{4}(2+\Delta)}3^{-\frac{3}{4}(2+\Delta)}
\left(12m^2-19m+4\right)\big]+2\eta^2 864^{-\frac{2+\Delta}{4}}
\bigg\}\,G^{\frac{2-\Delta}{4}}+36\left(m-1\right)6^{-\frac{m(\eta-1)}{2(\eta-2)}}\left[\Delta-\left(m+1\right)\right]\left[\frac{1}{m^3\left(m-1\right)}\right]^{\frac{3m(\eta-1)}{2(\eta-2)}}\\[2mm]
\nonumber
&&\hspace{-5mm}\times\,48^{-\frac{m(\eta-1)}{\eta-2}}\,\eta\left(\eta-2\right)^2\rho_0\left(\frac{2+\Delta}{2}\right)G^{\frac{3m(\eta-1)}{2(\eta-2)}}
\Bigg\}\frac{1}{3\left(2+\Delta\right)\left[\Delta-\left(m+1\right)\right]\left[4+3m\left(\eta-1\right)-2\eta\right]\left[2-\eta+m\eta\left(7\eta-8\right)\right]}\,+\,\eta T\,.
\label{GquSol2}
\end{eqnarray}

\subsection{Cosmological parameters}

In the following section, we present an analysis of several relevant cosmological parameters, which provide further insight into the dynamical behavior and observational viability of the model.

\subsubsection{Model I}
Using the definition~\eqref{defom} alongside Eqs.~\eqref{rhogt},~\eqref{pGT},~\eqref{constra}
and~\eqref{secf}, we get the following expression of the EoS parameter for Model I:
\begin{eqnarray}
\nonumber
&&\hspace{-4mm}\omega_{GT}\,=\,-\frac{1}{3GmC\left(m+3\right)}
\Bigg\{2^{\frac{6+m+\Delta}{4}}\,3^{\frac{14+m+3\Delta}{4}}\hspace{0.2mm}\eta_1 \rho_0^{\frac{1}{2}} 
\left[24m^3\left(m-1\right)\right]^{-\frac{3m}{8}}
\left(-5m^2+7m-2\right)G^{\frac{3m}{8}}\\[2mm]
\nonumber
&&\hspace{-4mm}+\,2^{\frac{6+\Delta}{4}}3^{\frac{14+m+3\Delta}{4}}19^{\frac{m}{8}}\hspace{0.2mm}
\eta_1\rho_0^{\frac{1}{2}}\left[114m^3\left(m-1\right)\right]^{-\frac{3m}{8}}\left(3m^2-11m+8\right)
G^{\frac{3m}{8}}
+24^{-\left(2+\Delta\right)}C\left[-\frac{m}{144\left[m^3(m-1)\right]^{\frac{1}{4}}}\right]^{-(2+\Delta)}\\[2mm]
\nonumber
&&\hspace{-4mm}\times\left[3m^2+\left(2+\Delta\right)\left(m+3\right)\right]
G^{\frac{2-\Delta}{4}}
+2^{\frac{20-9m+2\Delta}{8}}\hspace{0.2mm}3^{\frac{28-3m+6\Delta}{8}}
\eta_1\rho_0^{\frac{1}{2}}\left(m^2+2m-3\right)\left[\frac{G}{m^3(m-1)}\right]^{\frac{3m}{8}}+2^{\frac{22+\Delta}{4}}\hspace{0.2mm}3^{\frac{7}{2}+\frac{3\Delta}{4}}\eta_2\left(m-1\right)\\[2mm]
&&\hspace{-4mm}+\, 8^{-\left(2+\Delta\right)}C\left[-\frac{m}{48\left[m^3(m-1)\right]^{\frac{1}{4}}}\right]^{-(2+\Delta)}\left(5m-12\right)
G^{\frac{2-\Delta}{4}}\Bigg\}\left[-\frac{m\hspace{0.2mm}G^{\frac{1}{4}}}{6\left[m^3(m-1)\right]^{\frac{1}{4}}}\right]^{2+\Delta}\,.
\label{om1}
\end{eqnarray}

The corresponding deceleration parameter, derived from the definition~\eqref{qeq}, reads
\begin{eqnarray}
\nonumber
&&\hspace{-4mm}q\,=\,\frac{1}{2}-\frac{1}{3GmC\left(m+3\right)}
\Bigg\{2^{\frac{6+m+\Delta}{4}}\,3^{\frac{14+m+3\Delta}{4}}\hspace{0.2mm}\eta_1 \rho_0^{\frac{1}{2}} 
\left[24m^3\left(m-1\right)\right]^{-\frac{3m}{8}}
\left(-5m^2+7m-2\right)G^{\frac{3m}{8}}\\[2mm]
\nonumber
&&\hspace{-4mm}+\,2^{\frac{6+\Delta}{4}}3^{\frac{14+m+3\Delta}{4}}19^{\frac{m}{8}}\hspace{0.2mm}
\eta_1\rho_0^{\frac{1}{2}}\left[114m^3\left(m-1\right)\right]^{-\frac{3m}{8}}\left(3m^2-11m+8\right)
G^{\frac{3m}{8}}
+24^{-\left(2+\Delta\right)}C\left[-\frac{m}{144\left[m^3(m-1)\right]^{\frac{1}{4}}}\right]^{-(2+\Delta)}\\[2mm]
\nonumber
&&\hspace{-4mm}\times\left[3m^2+\left(2+\Delta\right)\left(m+3\right)\right]
G^{\frac{2-\Delta}{4}}
+2^{\frac{20-9m+2\Delta}{8}}\hspace{0.2mm}3^{\frac{28-3m+6\Delta}{8}}
\eta_1\rho_0^{\frac{1}{2}}\left(m^2+2m-3\right)\left[\frac{G}{m^3(m-1)}\right]^{\frac{3m}{8}}+2^{\frac{22+\Delta}{4}}\hspace{0.2mm}3^{\frac{7}{2}+\frac{3\Delta}{4}}\eta_2\left(m-1\right)\\[2mm]
&&\hspace{-4mm}+\, 8^{-\left(2+\Delta\right)}C\left[-\frac{m}{48\left[m^3(m-1)\right]^{\frac{1}{4}}}\right]^{-(2+\Delta)}\left(5m-12\right)
G^{\frac{2-\Delta}{4}}\Bigg\}\left[-\frac{m\hspace{0.2mm}G^{\frac{1}{4}}}{6\left[m^3(m-1)\right]^{\frac{1}{4}}}\right]^{2+\Delta}\,.
\label{qeq2}
\end{eqnarray}

In order to analyze the stability of the model under small perturbations, we now focus on the evolution of the squared sound speed $v_s^2$. By substituting Eqs.~\eqref{rhogt},~\eqref{pGT}, and~\eqref{cont1} into Eq.~\eqref{vsq}, and making use of the definition~\eqref{f1} of $f(G,T)$, we obtain  
\begin{eqnarray}
\nonumber
&&v_s^2\,=\,-\frac{1}{3\hspace{0.2mm}G\hspace{0.2mm}m\hspace{0.2mm}C\left(\Delta-2\right)\left(m+3\right)}
\Bigg\{2^{\frac{2+\Delta}{4}}3^{\frac{m+3(6+\Delta)}{4}}19^{\frac{m}{8}}
\eta_1\hspace{0.2mm}\rho^{\frac{1}{2}}\left[114m^3\left(m-1\right)\right]^{-\frac{3m}{8}}\left[m\left(-3m^2+11m-8\right)\right]G^{\frac{3m}{8}}\\[2mm]
\nonumber
&&+\,2^{\frac{-9m+2(6+\Delta)}{8}}3^{-\frac{3\left[m-2\left(6+\Delta\right)\right]}{8}}\eta_1\hspace{0.2mm}\rho^{\frac{1}{2}}\left[\frac{G}{m^3(m-1)}\right]^{\frac{3m}{8}}\left[m\left(-m^2-2m+3\right)\right]
+2^{\frac{2+m+\Delta}{4}}\hspace{0.2mm}3^{\frac{m+3(6+\Delta)}{4}}
\eta_1\hspace{0.2mm}\rho^{\frac{1}{2}}\left[24m^3(m-1)\right]^{-\frac{3m}{8}}\\[2mm]
\nonumber
&&\times\left[m\left(5m^2-7m+2\right)\right]
G^{\frac{3m}{8}}
+24^{-(2+\Delta)}C\left[-\frac{m}{144\left[m^3(m-1)\right]^{\frac{1}{4}}}\right]^{-(2+\Delta)}\left[3m^2\left(\Delta-2\right)+\left(2+\Delta\right)\left(m-24\right)\right]
G^{\frac{2-\Delta}{4}}\\[2mm]
&&+\,2^{-(4+3\Delta)}
\left[-\frac{m}{48\left[m^3(m-1)\right]^{\frac{1}{4}}}\right]^{-(2+\Delta)}
\left[\left(\frac{2+\Delta}{2}\right)^2\left(m+3\right)
+\left(-5m+12\right)
\right]
G^{\frac{2-\Delta}{4}}\Bigg\}
\left[-\frac{m\hspace{0.2mm}G^{\frac{1}{4}}}{6\left[m^3(m-1)\right]^{\frac{1}{4}}}\right]^{2+\Delta}\,.
\label{vsq1}
\end{eqnarray}

\subsubsection{Model II}
Let us now extend the above analysis to Model II, as defined in Eq.~\eqref{f2}. For the equation-of-state parameter, we obtain the following expression:
\begin{eqnarray}
\nonumber
&&\omega_{GT}=-\frac{1}{3GmC\left(m-1\right)\left(\eta-2\right)
\left[3m\left(\eta-1\right)+2\left(2-\eta\right)\right]\left[m\left(7\eta-8\right)+2-\eta\right]}\Bigg\{3^{\frac{2+\Delta}{2}}
\left[-\frac{m\hspace{0.2mm}G^{\frac{1}{4}}}{6\left[m^3(m-1)\right]^{\frac{1}{4}}}\right]^{2+\Delta}\\[2mm]
\nonumber
&&\times\bigg\{48^{-\frac{m(\eta-1)}{\eta-2}}6^{\frac{2m(1-\eta)+(10+\Delta)(\eta-2)}{4(\eta-2)}}
\rho_0
\left(\frac{G}{m^3(m-1)}\right)^{\frac{3m(\eta-1)}{2(\eta-2)}}
\left[m\hspace{0.2mm}\eta\left(2\,\frac{\eta^3-6\eta^2+12\eta-8}{m}
+\,3m^3\left(7\eta^3-29\eta^2+38\eta-16\right)
\right.\right.\\[2mm]
\nonumber
&&+m^2\left(-59\eta^3+261\eta^2-372\eta+172\right)
+3m\left(19\eta^3-250\eta^2+142\eta-72\right)
-3\left(\eta-2\right)^2\left(7\eta-9\right)
\bigg)\bigg]+\,2^{\frac{1}{4}\left[10+\Delta-\frac{18m(\eta-1)}{\eta-2}\right]}\\[2mm]
\nonumber
&&\times\,
3^{\frac{1}{4}\left[14+\Delta-\frac{6m(\eta-1)}{\eta-2}\right]}
\rho_0\left(\frac{G}{m^3(m-1)}\right)^{\frac{3m(\eta-1)}{2(\eta-2)}}
\left[m\eta^2\bigg(2\,\frac{\eta^2-4\eta+4}{m}+3m^3\eta\left(7\eta-15\right)+m^2\left(-59\eta^2+143\eta-62\right)
\right.\\[2mm]
\nonumber
&&+\,3m\left(19\eta^2+36\right)
-21\eta^2+69\eta-54\bigg)\bigg] + 12^{-\frac{2+\Delta}{2}}C
\left[-\frac{1}{12}\left(\frac{m}{m-1}\right)^{\frac{1}{4}}\right]^{-\left(2+\Delta\right)}\big\{9m^4\left(7\eta^3-29\eta^2+38\eta-16\right)\\[2mm]
\nonumber
&&+3m^3\left[-52\eta^3+232\eta^2-314\eta+156+\Delta\left(7\eta^3-29\eta^2+48\eta-16\right)\right]+m^2\left[127\eta^3-633\eta^2+1020\eta-524-2\Delta\right.\\[4mm]
\nonumber
&&\times\left.\left(19\eta^3+129\eta-62\right)\right]
+m\left[-2\left(17\eta^3-107\eta^2+204\eta+92\right)+\Delta\left(17\eta^3-95\eta^2+168\eta-92\right)\right]+
8\left[(\eta^2-3\eta)\right.\\[2mm]
\nonumber
&&\left.\times\left(\Delta-2\right)-4\right]
\big\}\,G^\frac{2-\Delta}{4}+ 
2^{-\left(3+2\Delta\right)}3^{-\frac{3\left(2+\Delta\right)}{2}}C\eta^5m\left[-\frac{1}{72}\left(\frac{m}{m-1}\right)^{\frac{1}{4}}\right]^{-\left(2+\Delta\right)}\left(\eta-2\right)\left(\Delta+3m-2\right)\left(m-1\right)G^{\frac{2-\Delta}{4}}\\[2mm]
&&+\,2^{\frac{1}{4}\left[22+\Delta-\frac{18m(\eta-1)}{\eta-2}\right]}
3^{\frac{1}{4}\left[18+\Delta-\frac{6m(\eta-1)}{\eta-2}\right]}m^3\eta^2
\rho_0\left[\frac{1}{m^3\left(m-1\right)}\right]^{\frac{3m(\eta-1)}{2(\eta-2)}}
G^{\frac{3m(\eta-1)}{2(\eta-2)}}\bigg\}\Bigg\}\,.
\label{EoS2}
\end{eqnarray}

In turn, the deceleration parameter reads
\begin{eqnarray}
\nonumber
&&q=\frac{1}{2}-\frac{1}{GmC\left(m-1\right)\left(\eta-2\right)
\left[3m\left(\eta-1\right)+2\left(2-\eta\right)\right]\left[m\left(7\eta-8\right)+2-\eta\right]}\Bigg\{3^{\frac{2+\Delta}{2}}
\left[-\frac{m\hspace{0.2mm}G^{\frac{1}{4}}}{6\left[m^3(m-1)\right]^{\frac{1}{4}}}\right]^{2+\Delta}\\[2mm]
\nonumber
&&\times\bigg\{48^{-\frac{m(\eta-1)}{\eta-2}}6^{\frac{2m(1-\eta)+(10+\Delta)(\eta-2)}{4(\eta-2)}}
\rho_0
\left(\frac{G}{m^3(m-1)}\right)^{\frac{3m(\eta-1)}{2(\eta-2)}}
\left[m\hspace{0.2mm}\eta\left(2\,\frac{\eta^3-6\eta^2+12\eta-8}{m}
+\,3m^3\left(7\eta^3-29\eta^2+38\eta-16\right)
\right.\right.\\[2mm]
\nonumber
&&+m^2\left(-59\eta^3+261\eta^2-372\eta+172\right)
+3m\left(19\eta^3-250\eta^2+142\eta-72\right)
-3\left(\eta-2\right)^2\left(7\eta-9\right)
\bigg)\bigg]+\,2^{\frac{1}{4}\left[10+\Delta-\frac{18m(\eta-1)}{\eta-2}\right]}\\[2mm]
\nonumber
&&\times\,
3^{\frac{1}{4}\left[14+\Delta-\frac{6m(\eta-1)}{\eta-2}\right]}
\rho_0\left(\frac{G}{m^3(m-1)}\right)^{\frac{3m(\eta-1)}{2(\eta-2)}}
\left[m\eta^2\bigg(2\,\frac{\eta^2-4\eta+4}{m}+3m^3\eta\left(7\eta-15\right)+m^2\left(-59\eta^2+143\eta-62\right)
\right.\\[2mm]
\nonumber
&&+\,3m\left(19\eta^2+36\right)
-21\eta^2+69\eta-54\bigg)\bigg] + 12^{-\frac{2+\Delta}{2}}C
\left[-\frac{1}{12}\left(\frac{m}{m-1}\right)^{\frac{1}{4}}\right]^{-\left(2+\Delta\right)}\big\{9m^4\left(7\eta^3-29\eta^2+38\eta-16\right)\\[2mm]
\nonumber
&&+3m^3\left[-52\eta^3+232\eta^2-314\eta+156+\Delta\left(7\eta^3-29\eta^2+48\eta-16\right)\right]+m^2\left[127\eta^3-633\eta^2+1020\eta-524-2\Delta\right.\\[4mm]
\nonumber
&&\times\left.\left(19\eta^3+129\eta-62\right)\right]
+m\left[-2\left(17\eta^3-107\eta^2+204\eta+92\right)+\Delta\left(17\eta^3-95\eta^2+168\eta-92\right)\right]+
8\left[(\eta^2-3\eta)\right.\\[2mm]
\nonumber
&&\left.\times\left(\Delta-2\right)-4\right]
\big\}\,G^\frac{2-\Delta}{4}+ 
2^{-\left(3+2\Delta\right)}3^{-\frac{3\left(2+\Delta\right)}{2}}C\eta^5m\left[-\frac{1}{72}\left(\frac{m}{m-1}\right)^{\frac{1}{4}}\right]^{-\left(2+\Delta\right)}\left(\eta-2\right)\left(\Delta+3m-2\right)\left(m-1\right)G^{\frac{2-\Delta}{4}}\\[2mm]
&&+\,2^{\frac{1}{4}\left[22+\Delta-\frac{18m(\eta-1)}{\eta-2}\right]}
3^{\frac{1}{4}\left[18+\Delta-\frac{6m(\eta-1)}{\eta-2}\right]}m^3\eta^2
\rho_0\left[\frac{1}{m^3\left(m-1\right)}\right]^{\frac{3m(\eta-1)}{2(\eta-2)}}
G^{\frac{3m(\eta-1)}{2(\eta-2)}}\bigg\}\Bigg\}\,.
\label{dec2}
\end{eqnarray}

Concerning the squared sound speed $v_s^2$, we finally acquire
\begin{eqnarray}
\nonumber
v_s^2&=&\frac{\left(\eta -1\right) \eta  \left(2 \eta -1\right) m\hspace{0.2mm} \rho_0\hspace{0.2mm} G^{\frac{\Delta -2}{4}} \left(-\sqrt[4]{\frac{m}{m-1}}\right)^{\Delta -2} \left[\frac{\left(m-1\right) m^3}{G}\right]^{-\frac{3 \left(\eta -1\right) m}{2 (\eta -2)}} 2^{-\frac{3 \Delta }{4}-\frac{9 (\eta -1) m}{2 (\eta -2)}+\frac{5}{2}}\, 3^{\frac{1}{4} \left[-\Delta -\frac{6 \left(\eta -1\right) m}{\eta -2}+6\right]}}{C \left(\Delta -2\right) \left(\eta -2\right)^2}\\[2mm]
&&-\,\frac{\Delta+3m-2}{3m}\,.
\label{vs2}
\end{eqnarray}

\acknowledgments 
This research is supported by the postdoctoral fellowship program of the University of Lleida. The author gratefully acknowledges the contribution of the LISA Cosmology Working Group (CosWG), as well as support from the COST Actions CA21136 - \textit{Addressing observational tensions in cosmology with systematics and fundamental physics (CosmoVerse)} - CA23130, \textit{Bridging high and low energies in search of quantum gravity (BridgeQG)} and CA21106 -  \textit{COSMIC WISPers in the Dark Universe: Theory, astrophysics and experiments (CosmicWISPers)}.

\end{document}